\documentclass[10pt, conference, compsocconf]{IEEEtran}
\ifCLASSINFOpdf
   \usepackage[pdftex]{graphicx}
\else
   \usepackage[dvips]{graphicx}
\fi

\usepackage{comment}

\usepackage{numprint}
\npthousandsep{,}
\usepackage{color}

\usepackage{listings}

\usepackage{ifthen}
\newboolean{longVersion}
\setboolean{longVersion}{true}

\usepackage[tight,footnotesize]{subfigure}
\begin{document}
%
\title{Percolation Computation in Complex Networks}

\author{\IEEEauthorblockN{Fergal Reid}
\IEEEauthorblockA{Clique Research Cluster\\
Complex Adaptive Systems Laboratory\\
University College Dublin\\
Email: fergal.reid@gmail.com}
\and
\IEEEauthorblockN{Aaron McDaid}
\IEEEauthorblockA{Clique Research Cluster\\
Complex Adaptive Systems Laboratory\\
University College Dublin\\
Dublin 4, Ireland\\
}
\and
\IEEEauthorblockN{Neil Hurley}
\IEEEauthorblockA{Clique Research Cluster\\
Complex Adaptive Systems Laboratory\\
University College Dublin\\
Dublin 4, Ireland\\
}

}

\maketitle

\begin{abstract}
$K$-clique percolation is an overlapping community finding algorithm which extracts particular structures, comprised of overlapping cliques, from complex networks.
While it is conceptually straightforward, and can be elegantly expressed using clique graphs, certain aspects of $k$-clique percolation are computationally challenging in practice.
In this paper we investigate aspects of empirical social networks, such as the large numbers of overlapping maximal cliques contained within them, that make clique percolation, and clique graph representations, computationally expensive.
We motivate a simple algorithm to conduct clique percolation, and investigate its performance compared to current best-in-class algorithms.
We present improvements to this algorithm\footnote{Source code available: http://sites.google.com/site/CliquePercComp}, which allow us to perform $k$-clique percolation on much larger empirical datasets.
Our approaches perform much better than existing algorithms on networks exhibiting pervasively overlapping community structure, especially for higher values of $k$.
However, clique percolation remains a hard computational problem; current algorithms still scale worse than some other overlapping community finding algorithms.
\end{abstract}


\IEEEpeerreviewmaketitle

\section{Introduction}
One particular type of percolation, important in the study of complex networks, studied by Palla et al. \cite{palla2005uncovering} is \textit{$k$-clique percolation}, which Fortunato's review \cite{fortunato2009community} of community detection methods describes as the most popular overlapping community detection method.
A \emph{clique} is a group of nodes in a network, such that every node is connected to each other node.
Palla et al. argue that percolated $k$-cliques -- groups of cliques of size $k$, that are connected together by cliques of size ($k$-1) -- are important structures in complex networks, and a good way to find community structure.
They discuss $k$-clique percolation in a variety of application contexts, including authorship networks, word association networks, and certain biological contexts, for example, claiming that percolated $k$-cliques in protein-protein interaction networks correspond to functional units of protein structure.
Additionally, Palla, Barabasi, and Vicsek \cite{palla2007quantifying} apply $k$-clique percolation to the study of social dynamics, in mobile telecoms social network datasets.

More generally, the work of Evans \cite{evans2010clique} motivates `clique graphs' as interesting constructions to use when studying structure in networks and finding communities.
A `clique graph' of a specific source network is formed by representing each clique in the source network by a node in the `clique graph'. 
Pairs of nodes in the clique graph are then connected by edges where the corresponding pair of cliques overlap in the source network. 
Clique graphs have many interesting properties, not least that $k$-clique percolation can be expressed as a simple thresholding on the clique graph.
Clique graph construction and $k$-clique percolation are thus linked. 
Evans argues that different types of partitioning and thresholding of the clique graph are interesting to examine for the purposes of overlapping community detection, and also to examine structure in complex networks generally.

However, in practice these approaches to find network structure present computational challenges.
In a naive algorithmic approach to $k$-clique percolation, one algorithm to percolate $k$-cliques -- for example as described in the review of Fortunato, citing the work of Everett and Borgatti \cite{everett1998analyzing} -- is ``\emph{In order to find $k$-clique communities, one searches first for maximal cliques. Then a clique-clique overlap matrix $O$ is built, which is an $n_c \times n_c$ matrix, $n_c$ being the number of cliques; $O_{ij}$ is the number of vertices shared by cliques $i$ and $j$. To find $k$-cliques, one needs simply to keep the entries of $O$ which are larger than or equal to $k - 1$, set the others to zero and find the connected components of the resulting matrix.}''
This simple algorithm, which essentially finds $k$-clique percolations by finding connected components in a certain construction of the clique graph, is quadratic in the number of maximal cliques in the graph, by virtue of the way that it builds the complete clique-clique overlap matrix.
Furthermore, the number of cliques, and maximal cliques, in an empirical network may be \emph{very} large, as we will discuss.

Some work has presented faster computational methods of obtaining percolated $k$-clique structures.
The authors of the original Palla et al. paper have provided a faster CFinder \cite{adamcsek2006cfinder} implementation to find percolated $k$-cliques.
More recent work has been done by Kumpula et al. with their `Sequential Clique Percolation' algorithm (SCP) \cite{kumpula2008sequential} which further improves efficiency. 
However, these improved methods often perform poorly on networks with the kind of pervasively overlapping community structure we see in many real world social networks -- an area of increasing interest in the applied study of community structure -- and particularly poorly when performing percolation with high values of $k$.
In this work, we consider several computational aspects of the problem of percolating structures in large complex networks.
We focus our discussion on $k$-clique percolation, but the techniques we describe could be applied to the computation of many similar percolation problems.

\ifthenelse {\boolean{longVersion}} {
We note that there are many other overlapping community finding methods, some of which are much more scalable than clique percolation; however in this paper we focus on clique percolation, due to its popularity. We also hope that $k$-clique percolation will serve as a case study for computational techniques that could later be used to improve the scalability of algorithms for other types of percolation.
}

\subsection{Structure of the paper}
In Section \ref{SECkcliquePercolation} we define clique percolation. We also discuss clique graphs,
which are a useful conceptual tool to understand the problem and to gain insight into various algorithms.
We show how, even though cliques are indicative of community structure, there can be many more 
cliques than communities.

Algorithms typically fall into one of two classes: either they are based on finding the $k$-cliques
or else they limit themselves to the \emph{maximal} cliques with at least $k$ nodes in them. 
We explain how these are equivalent and discuss an example of each kind from the literature.

In Section \ref{SECchallenges} we discuss some challenges that face all the examples in
the literature and which are also faced by our algorithms. The number of cliques
can be very large, as can the size of the clique graph.
Also we explain why, while it is tempting to attempt to represent a community of cliques
merely as the union of the nodes in its constituent cliques, such a method
will lead to an incorrect algorithm.

In Section \ref{SECourapproaches}, we define and motivate our algorithms and evaluate the
speed of them against SCP on a variety of synthetic networks and on empirical social networks.

\ifthenelse {\boolean{longVersion}} {
In Section \ref{SECotherApproaches}, we briefly mention other approaches to this problem that we considered and other overlapping community finding algorithms of interest.
} 

\section{$K$-clique Percolation}
\label{SECkcliquePercolation}

\ifthenelse {\boolean{longVersion}} {
We now give a description of $k$-clique percolation, constraining ourselves to computationally relevant properties.
For a detailed introduction to the method, and arguments motivating it as a tool for overlapping community finding, consult Palla et al. \cite{palla2005uncovering}.
}

In $k$-clique percolation, if two cliques of size $k$ share \mbox{$k$-1} nodes, we say that these two cliques percolate into each other. 
The maximal sets of cliques, which satisfy the property that every clique in the set is reachable from every other clique in the set, through a path connecting percolating pairs, form the communities output by k-clique percolation.
These communities are then typically output as the sets of nodes contained within each set of cliques -- and may overlap.
\ifthenelse {\boolean{longVersion}} {
Typically, for a given community found by $k$-clique-percolation, a particular application requires these sets of nodes as the representation of the communities found.
However, as we will see later, in order to
maintain a correct understanding of clique percolation it is best to think about $k$-clique communities as sets of overlapping cliques, rather than the set of nodes that each set of percolated cliques contains.
}

\subsection{Clique Graphs}
A clique graph is formed from a source network by creating a node in the clique graph for each clique in the source network and joining two nodes in the clique graph if their corresponding cliques in the source network share nodes, or `overlap' -- see Figure \ref{cliqueGraphSimple} for a visualisation.
\ifthenelse {\boolean{longVersion}} {
This is a simple visualisation of a small clique graph extracted from a simple network; such relationships are difficult to visualise due to the heavily overlapping nature of cliques in networks; this is especially true as the value of $k$ used is increased.
}

The edges in a clique graph may be weighted, corresponding to the overlap between the cliques.  
Thresholding such a graph -- removing the nodes corresponding to cliques of size less than some value $k$, and removing edges with weight lower than $(k - 1)$ -- yields a graph in which the connected components are the cliques that would have percolated into each other, in $k$-clique percolation. 
Figures \ref{minimalCliqueVisualisation}a, \ref{minimalCliqueVisualisation}b, and \ref{minimalCliqueVisualisation}c illustrate this process.
As both thresholding the edges of a graph and calculating the connected components of a graph are computationally inexpensive, it is inexpensive to calculate the $k$-clique percolations, given the clique graph.

\ifthenelse {\boolean{longVersion}} {
The clique graph is a useful conceptual tool to understand the definition of $k$-clique percolation, and to understand the particular algorithms that have been used to recover these structures.
Evans \cite{evans2010clique} discusses some other general advantages of using clique graphs: ``The advantage is that there are many well established methods for analysing the properties of vertices of a graph and for the price of a simple transformation, these can be applied to obtain the same information about the cliques'', and goes on to state that ``In terms of computational efficiency, the clique graphs are generally bigger [than the source network] but by how much depends on the detailed structure of the graph. The speed savings of a good fast vertex partitioning algorithm [...] may compensate for the larger size of the clique graph.''
}

However, while clique graphs are an elegant construction, and a useful tool for reasoning about clique percolation, there are computational obstacles to their use on many empirical networks, as we shall see; in practice, the clique graph often takes a very long time to construct. 
\ifthenelse {\boolean{longVersion}} {

We will also later see that, in the specific case of $k$-clique percolation, it is in fact not necessary to compute the entire clique graph in order to compute the percolated communities.
}
{
We will also later see that it is in fact \emph{not} necessary to compute the entire clique graph in order to compute the percolated communities.
}

\ifthenelse {\boolean{longVersion}} {
\begin{figure}[!htb]
\begin{center}
\caption{This illustration shows a very simple network, and its clique graph.  We represent the source network with circular nodes. We represent the maximal cliques in this network as squares, placing a square at the center of each 4-clique.  The corresponding maximal clique graph, with threshold of overlap $3$, is shown below. The connected components in this graph -- there is only one in our example -- correspond to the cliques that percolate into each other -- this entire graph would be one $k$-clique percolation, for $k=4$.}
\vspace{5px}
    \includegraphics[width=50mm]{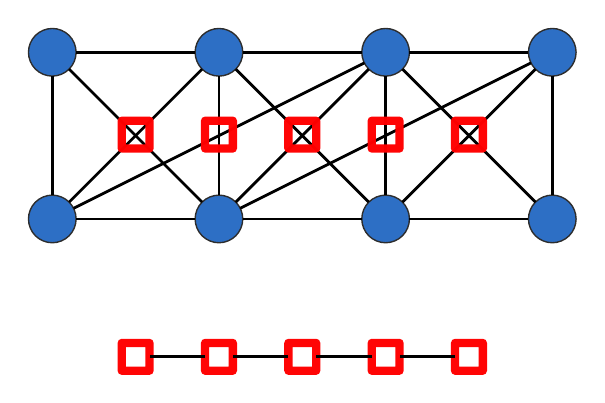}
\label{cliqueGraphSimple}
\vspace{-10px}
\end{center}
\end{figure}
}
{
\begin{figure}[!htb]
\begin{center}
    \subfigure[]{
        \includegraphics[width=41mm]{cliqueGraphSimple.pdf}
        \label{cliqueGraphSimple}
    }
    \subfigure[]{
        \includegraphics[width=41mm]{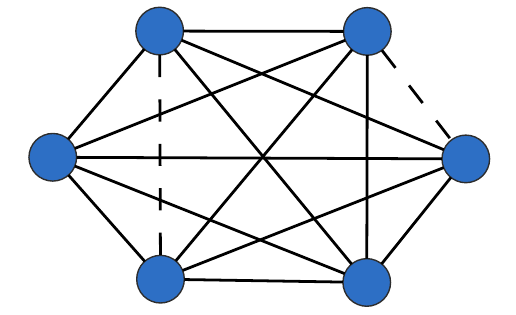}
        \label{moreCliquesThanComms}
    }
\caption{ (a) This illustration shows a very simple network, and its clique graph.  We represent the source network with circular nodes. We represent the maximal cliques in this network as squares, placing a square at the center of each 4-clique.  The corresponding maximal clique graph, with threshold of overlap $3$, is shown below. The connected components in this graph -- there is only one in our example -- correspond to the cliques that percolate into each other -- this entire graph would be one $k$-clique percolation, for $k=4$. 
\newline(b) A maximal 6-clique. If the dashed edges are removed, this diagram shown now contains 4 maximal cliques. This counter-intuitive behaviour, whereby missing edges can increase the number of maximal cliques in a network, partly illustrates why a particular empirical network can have many more maximal cliques than it has edges, nodes, and communities.}
\label{}
\end{center}
\vspace{-12px}
\end{figure}
}

\subsection{Cliques vs. Maximal Cliques}
Throughout this paper, we discuss $k$-clique percolation, with reference to two definitions of `clique':
A `clique' is a fully connected sub-graph -- a set of nodes all of which are connected to each other.
A `maximal clique' is a clique that is contained in no larger clique.
Every maximal clique is a clique, by definition, but the opposite does not hold.
Thus there are always more cliques than maximal cliques.

\ifthenelse {\boolean{longVersion}} {
In fact, as has been previously \cite{palla2005uncovering}\cite{kumpula2008sequential} pointed out, any maximal clique of size $n$ will always have ${{n}\choose{k}}$ cliques of size $k$, for each $k<n$, within it.
Note, however, that these contained cliques will not be distinct across maximal cliques.
Nonetheless, by definition, each maximal clique will have at least one clique within it that is not contained in any other clique, and usually many more.
Thus, the number of cliques in any non-trivial network is always greater than the number of maximal cliques.
This is an important consideration, as some algorithms have complexity in terms of cliques, and others in terms of maximal cliques.
}

It is important to realise that cliques of size $k$ will always percolate within a larger clique of size greater than $k$; thus a maximal clique of size greater than $k$ contains a $k$-clique percolation (which may also extend outside it). Further, any $k$-clique sharing $k$-1 nodes with this maximal clique will be part of the same percolation.
Hence, clique percolation can be equivalently defined in these two different ways; one definition is based on a clique graph
of every clique with exactly $k$ nodes, while another definition is based on a maximal-clique graph of
every maximal clique of at least $k$ nodes.
Various algorithmic implementations make use of each of these definitions, with consequences for computational performance.

\subsection{More cliques than communities}

The large numbers of maximal cliques found in the \mbox{`Facebook 100'} networks \cite{traud2008community} shown in Table \ref{numbersOfCliques} -- much greater than the number of nodes in each network -- may be surprising, given the sociological interpretation of cliques as a community structure \cite{luce1949method}; if a clique corresponds to the idea of a community, we may wonder why social networks contain so many more maximal cliques than nodes.
However, in empirical social networks, we often observe a situation where a large maximal clique may overlap heavily with many other smaller maximal cliques that differ from it by only a few nodes.
It is also the case, that if edges are randomly deleted from a single maximal clique, this maximal clique will often be replaced by many smaller maximal cliques.
A visualisation of this is shown in Figure \ref{moreCliquesThanComms}, where a single maximal clique, after having 2 edges removed, turns into 4 maximal cliques.
\ifthenelse {\boolean{longVersion}} {
Thus, in a social network setting, if not every friendship is present between a strong community of individuals, many maximal cliques may emerge within this single `community' of nodes.
This goes some way to intuitively explaining the large numbers of maximal cliques seen in datasets from on-line social networks. 
}

\ifthenelse {\boolean{longVersion}} {
\begin{figure}[!htb]
\begin{center}
\caption{A maximal 6-clique. If the dashed edges are removed, this diagram shown now contains 4 maximal cliques. This counter-intuitive behaviour, whereby missing edges can increase the number of maximal cliques in a network, partly illustrates why a particular empirical network can have many more maximal cliques than it has edges, nodes, and communities.}
    \includegraphics[width=50mm]{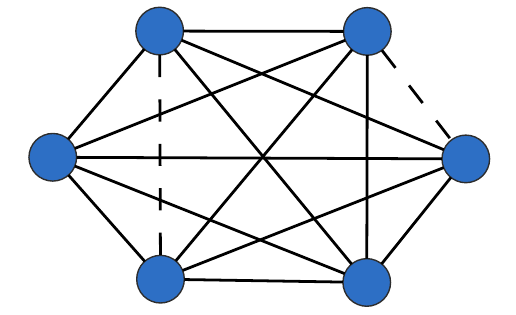}
\label{moreCliquesThanComms}
\end{center}
\end{figure}
}

\subsection{Previous approaches}
\subsubsection{CFinder}
The original CFinder  implementation, as described by Palla et al. in supplementary material \cite{palla2005uncovering}, follows the method of Everett and Borgatti, finding all \emph{maximal} cliques, by a custom method, and generating the overlap matrix between these cliques.
Once this overlap matrix is found, percolations for all values of $k$ can then easily be calculated.
However, building the full overlap matrix -- equivalent to the full clique graph -- naively requires $O(n_c^2)$ clique-clique comparisons, in $n_c$ the number of maximal cliques.
Thus, CFinder performs poorly on the empirical networks we study.

\subsubsection{SCP}
SCP \cite{kumpula2008sequential}, on the other hand, works off a custom clique finding algorithm to find, not all the \emph{maximal} cliques, but all cliques of size $k$.
For each clique of size $k$, or `$k$-clique', it is trivial to find the cliques of size $k$-1 inside it.
SCP is based on a bipartite graph with the $k$-cliques on one side and the \mbox{($k$-1)}-cliques on the other side.
A ($k$-1)-clique is linked to every \mbox{$k$-clique} containing it.
The connected components in this bipartite network correspond to the clique percolation communities.
The idea is that two ($k$-1)-cliques percolate into each other if they are both members of the same $k$-clique,
and this is equivalent to the conventional definition of two $k$-cliques being linked if they 
share $k$-1 nodes.

While SCP consistently outperforms CFinder \cite{kumpula2008sequential}, the important term for the performance of SCP on a network is thus not the number of maximal cliques of size at least $k$, but the number of cliques of size $k$-1.
As we discuss in our experimental work, empirical social networks often contain maximal cliques of substantial size; given that such large maximal cliques contain very many smaller $k$-cliques, the fact that SCP has complexity in terms of the number of cliques of size $k$-1 often leads to poor performance in practice.
We note that, as stated by Kumpula et al., SCP is most efficient for low sizes of $k$, and performs poorly as the value of $k$ increases.

\section{Computational Challenges}
\label{SECchallenges}

In this section, we look at some of the challenges that face any $k$-clique percolation algorithm.
We discuss our specific solutions and algorithms in the next section.

\subsection{Quantity of cliques}
To understand why $k$-clique percolation is so computationally expensive on certain types of online social networks, we first quantify the numbers of maximal cliques in these empirical networks.
The various algorithms deal with the
edges in the clique graph differently, but each algorithm requires all the (maximal) cliques be found first.

\ifthenelse {\boolean{longVersion}} {
As shown in Table \ref{numbersOfCliquesAll}, even networks that have small numbers of nodes and edges, can have very large numbers of maximal cliques.  
It is worth noting that the worst case networks, in terms of the ratio of maximal cliques to nodes in the network, are online social networks; such datasets, for example the Facebook networks, feature highly overlapping community structure and have very large numbers of maximal cliques.
}

Our primary interest, in this work, is in clique percolation in the challenging domain of modern online social networks, which exhibit pervasive overlap \cite{ahn2010link} \cite{reid2011partitioning}.
The `Facebook 100' dataset of Traud et al. \cite{traud2008community} provides us with a range of similar empirical networks of differing size.
We focus on the smallest 10 of these networks -- shown in Table \ref{numbersOfCliques} -- which capture the general topological features, in terms of densely overlapping community structure, but are also small enough to benchmark existing algorithms efficiently on.
There are very large numbers of maximal cliques in these networks, many more than the number of nodes or edges.

\ifthenelse {\boolean{longVersion}} {

\begin{table}[ht]
\caption{The vast number of maximal cliques (size$\ge$4) present in many empirical networks. Of particular note are the Facebook and Twitter datasets, with many more maximal cliques than nodes or edges. Datasets from SNAP \cite{SNAP} and other sources; see \cite{reid2011partitioning} for details.}
\begin{center}
\begin{tabular}{rlrp{1.0cm}p{1.0cm}}
  \hline
Network & Nodes & Edges & Maximal\par Cliques & Largest\par Clique  \\
  \hline
\hline
Email-Enron  & \numprint{36692} & \numprint{367662} & \numprint{205712} & 20  \\
Email-EuAll  & \numprint{265009} & \numprint{420045} & \numprint{93267} & 16  \\
Mobile1 & \numprint{10001} &\numprint{48556} & \numprint{1550} &10  \\  
Mobile2 & \numprint{10001} & \numprint{91930} &\numprint{3538} &10  \\  
Mobile3 & \numprint{10001} & \numprint{88714} &\numprint{951} & 9\\ 
\hline
Facebook-caltech  & \numprint{769} & \numprint{16656} & \numprint{31745} & 20 \\
Facebook-princeton  & \numprint{6596} & \numprint{293320} & \numprint{1286678} & 34\\
Facebook-georgetown  & \numprint{9414}  & \numprint{425638} &\numprint{1440853}  & 33 \\
Twitter1  & \numprint{2001} & \numprint{47000} & \numprint{23570} &12 \\ 
Twitter2  & \numprint{2001} & \numprint{71264}  & \numprint{554489} &27 \\ 
Twitter3  & \numprint{2001} & \numprint{48914} & \numprint{130399} &22\\ 
Slashdot0811  & \numprint{77360} & \numprint{516575} & \numprint{441941} & 26 \\
\hline
Collab-AstroPhysics  & \numprint{18771} &\numprint{396160} & \numprint{27997} & 57 \\
Collab-CondMat  & \numprint{23133} & \numprint{186936} & \numprint{8824} & 26 \\
Collab-HighEnergy  & \numprint{9875} & \numprint{237010} & \numprint{2636} & 32 \\
Cite-HighEnergy  & \numprint{27769} & \numprint{421578} & \numprint{419942} & 23 \\
\hline
Amazon0302  & \numprint{262111} &\numprint{1234877} & \numprint{117054} & 7 \\
Epinions  & \numprint{75879} & \numprint{841372} & \numprint{1596598} & 23 \\
\hline
Web-NotreDame  & \numprint{325729} & \numprint{1497134} & \numprint{130965} & 155 \\
Web-Stanford  & \numprint{281903} & \numprint{2312497}  & \numprint{774555} & 61 \\
Wiki-Vote  & \numprint{7115} & \numprint{103689} &\numprint{436629} & 17 \\
\hline
ProteinInteract-Collins & \numprint{1622} & \numprint{9070} & \numprint{4310} &33 \\
\hline
\end{tabular}
\label{numbersOfCliquesAll}
\end{center}
\end{table}
}

\begin{table}[ht]
\caption{The vast number of maximal cliques (size$>=$3) present, even in relatively small empirical on-line social networks. `CGBound' is the number of edges, in 1000s, we calculated in each $k=5$ maximal clique graph, after several weeks computation, and is an underestimate of the true value. (*Reed98 is the exact number, not an underestimate.)}
\begin{center}
\begin{tabular}{|r|l|r|p{1.0cm}|p{1.0cm}|p{1.0cm}|}
  \hline
Network & Nodes & Edges & Maximal\par Cliques & Largest\par Clique &CGBound\par (1000s)\\
  \hline
\hline
Caltech36 & \numprint{769} & \numprint{16656} & \numprint{32207} & 20 & \numprint{2475}\\ 
Reed98 & \numprint{962} & \numprint{18812} & \numprint{33991} & 16 & \numprint{1774}* \\ 
Simmons81 & \numprint{1518} & \numprint{32988} & \numprint{45538} & 19 & \numprint{2244}\\ 
Haverford76 & \numprint{1446} & \numprint{59589} & \numprint{475567} & 24& \numprint{127}\\ 
Swarthmore42 &\numprint{1659} & \numprint{61050} & \numprint{306542} & 20& \numprint{282}\\ 
USFCA72 & \numprint{2682} & \numprint{65252} & \numprint{108929} & 29 & \numprint{1128}\\ 
Mich67 & \numprint{3748} & \numprint{81903} & \numprint{154971} & 27 & \numprint{843}\\ 
Bowdoin47 & \numprint{2252} & \numprint{84387} & \numprint{331738} & 23 & \numprint{235}\\ 
Oberlin44 & \numprint{2920} & \numprint{89912} & \numprint{198803} & 22 & \numprint{644}\\ 
Amherst41 & \numprint{2235} & \numprint{90954} & \numprint{599430} & 21 & \numprint{79}\\ 
\hline
\end{tabular}
\label{numbersOfCliques}
\end{center}
\vspace{-8px}
\end{table}

\subsection{Clique Graph size}
From the large numbers of cliques present in these types of pervasively overlapping network, it can be seen that naively performing $O(n_c^2)$ intersection tests, in $n_c$ the number of cliques, in order to populate a clique-clique intersection matrix would be computationally prohibitive.

However, in addition, by attempting to measure the size of the maximal clique graph, we can see that even if we had a fast way to generate the full clique graph, it would be extremely unwieldy to work with, because of the vast number of edges in it.
\ifthenelse {\boolean{longVersion}} {
It is computationally expensive to obtain all the edges in the clique graph for these networks. Running a process to calculate maximal clique graph edges on one of the smallest Facebook networks (Caltech) for a period of 12 hours calculated in excess of 449,000 edges in the clique graph; this was the clique graph of maximal cliques of size greater than 7, with edges of intersection of at least 6 nodes. Similarly, looking at maximal cliques of size greater than 5 with intersections of 4 nodes, we generated over 
1,700,000 edges over a period of several days computation. 
We note that the total number of edges in the clique graph, in this small network with only 769 nodes, may even be much larger than this.
Here, we are calculating these clique graph edges by performing set-set intersection tests, for each clique, against the set of cliques with which it shares a node.
Table \ref{numbersOfCliques} shows an underestimate of the number of edges in each clique graph, formed after over one week of computation.
The number of actual edges present may be vastly greater, but merely dealing with clique graphs that are sometimes so many orders of magnitude larger than the original source network would be computationally prohibitive.
Even if there were a way to speed up these intersection tests, merely dealing with clique graphs that are so many orders of magnitude larger than the original source network fast becomes computationally prohibitive.
}
{It is computationally expensive to obtain all the edges in the clique graph for these networks, but running a process to calculate maximal clique graph edges, for a limited amount of time, allows us to calculate a lower bound on the numbers of clique graph edges.
Table \ref{numbersOfCliques} shows an underestimate of the number of edges in each clique graph, formed after over one week of computation.
The number of actual edges present may be vastly greater, but merely dealing with clique graphs that are sometimes so many orders of magnitude larger than the original source network would be computationally prohibitive.
}

These figures illustrate that methods of clique percolation, or indeed general graph analysis, which require the full clique graph to be built, are unscalable on these networks.
This motivates methods, such as those we present, which build only a subset of the clique graph.

\subsection{The necessity of dealing with individual cliques}
\ifthenelse {\boolean{longVersion}} {
Given the vast numbers of cliques present in empirical graphs, a second complication exists that makes clique percolation particularly computationally hard.
}

\begin{figure}[!htb]
\begin{center}
\caption{We cannot use an intermediate representation to store a percolating $k$-clique structure, without storing the individual cliques that comprise it: Consider the case where we store just the nodes and edges of the percolating 4-clique structure shown in the diagram.
Even if the highlighted dashed triangle was part of another 4-clique, with an additional node, not shown, this other 4-clique would \emph{not} percolate into the 4-clique-community comprising all the shown nodes; even though all three nodes and edges in the highlighted triangle are part of the percolated 4-clique community, the \emph{triangle} is not part of any of the constituent 4-cliques.
}
\vspace{4px}
    \includegraphics[width=50mm]{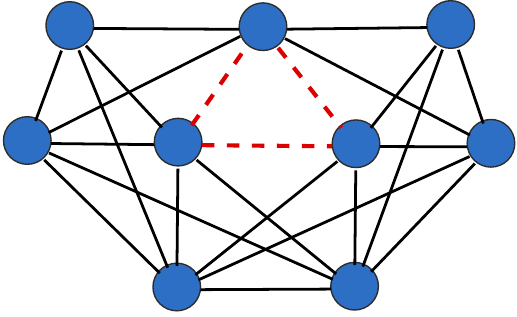}
\label{cliqueProblem2}
\vspace{-20px}
\end{center}
\end{figure}

As cliques are percolating, we would ideally discard the individual cliques that comprise a percolating structure, and instead describe that structure as a set of nodes and edges.
As there are so many more cliques than nodes or edges, this would increase efficiency.
However, we cannot do this, because it is not possible to tell, given a set of nodes, and the edges between them, whether a given clique percolates into this structure, without knowing the ($k$-1)-cliques that comprise this structure.
Counter-intuitively, a percolating $k$-clique structure can contain ($k$-1)-cliques that are not part of the $k$-cliques that yield it.
Figure \ref{cliqueProblem2} shows an example of this.
As such, it is not possible to test a candidate clique for percolation, against a percolating structure, without maintaining a list of the $k$, or $k$-1, cliques that make it up.
We believe this is a core computational issue in $k$-clique percolation.

\section{Our Algorithms}
\label{SECourapproaches}
In this section we describe two algorithms. 
Our algorithms initially obtain the maximal cliques in the graph, and then \mbox{attempt} to minimise the number of clique overlap tests that must be carried out to obtain $k$-clique communities.
\mbox{Algorithm 1} attempts to build a \emph{minimal spanning forest} over the maximal cliques, using a simple data structure to reduce unnecessary clique intersection tests.
Algorithm 2 uses a hierarchical data structure to further reduce the number of full intersection tests needed, further improving performance.

\subsubsection{Maximal Cliques}
Like CFinder we have chosen to build our algorithms around maximal cliques with at least $k$ nodes; this is in contrast to SCP, which uses $k$-cliques.
This is because large maximal cliques are often present in empirical data; these contain large numbers of $k$-cliques and thus it is often more efficient to work with the maximal cliques.
Figure \ref{CliqueSizeDistributions} shows the clique- and maximal-clique distributions on a typical Facebook network -- though there are many maximal cliques, there are many more $k$-cliques than there are maximal cliques of size greater than $k$, for all but the smallest values of $k$.

\subsubsection{Clique finding}
The Bron-Kerbosch clique finding algorithm \cite{bron1973finding} provides us with a fast way to find all maximal cliques in the network.
While finding all cliques, or maximal cliques, is computationally hard on an arbitrary graph, on the networks that we examine in practice, this method finds all clique extremely fast, such that clique-finding is a small part of the computational time in our algorithms.
Instead computation is dominated by comparing cliques against each other.

\subsection{Algorithmic framework}
\label{SECalgorithmTemplate}
Our goal is to find which cliques are in which connected components of
the thresholded clique graph.
We follow a procedure similar to the well known \cite{hopcroft1973algorithm} connected components algorithm.
We process one component at a time, by taking an arbitrary starting clique and identifying all the
cliques in the same connected component as it.
This process repeats until all cliques have been
assigned to a component.

\begin{figure*}[!htb]
\begin{center}
    \subfigure[The source network.]{
        \includegraphics[width=80mm]{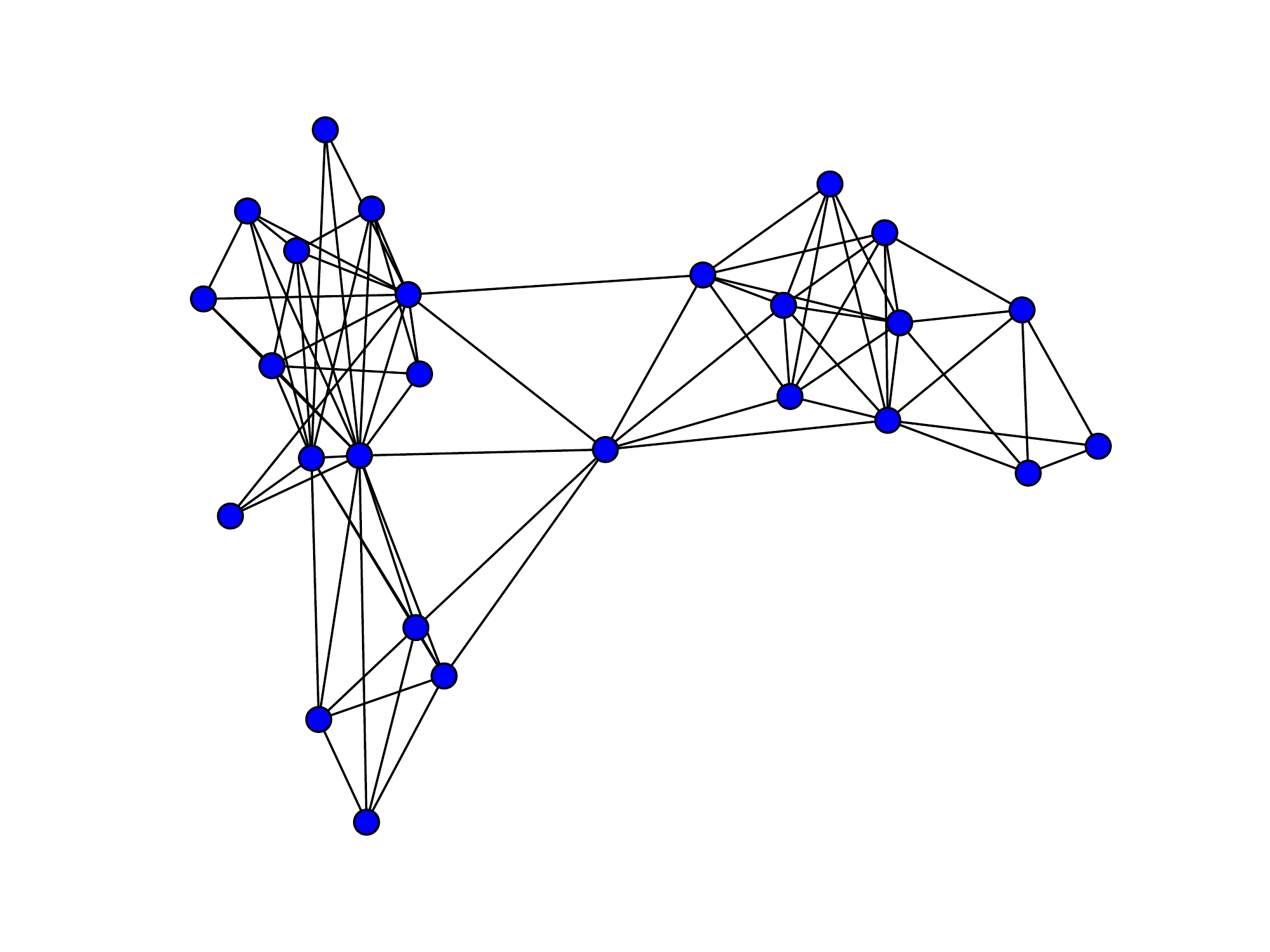}
        \label{1}
    }
    \subfigure[Maximal cliques of size at least 4, represented by a red box at their centroids.]{
        \includegraphics[width=80mm]{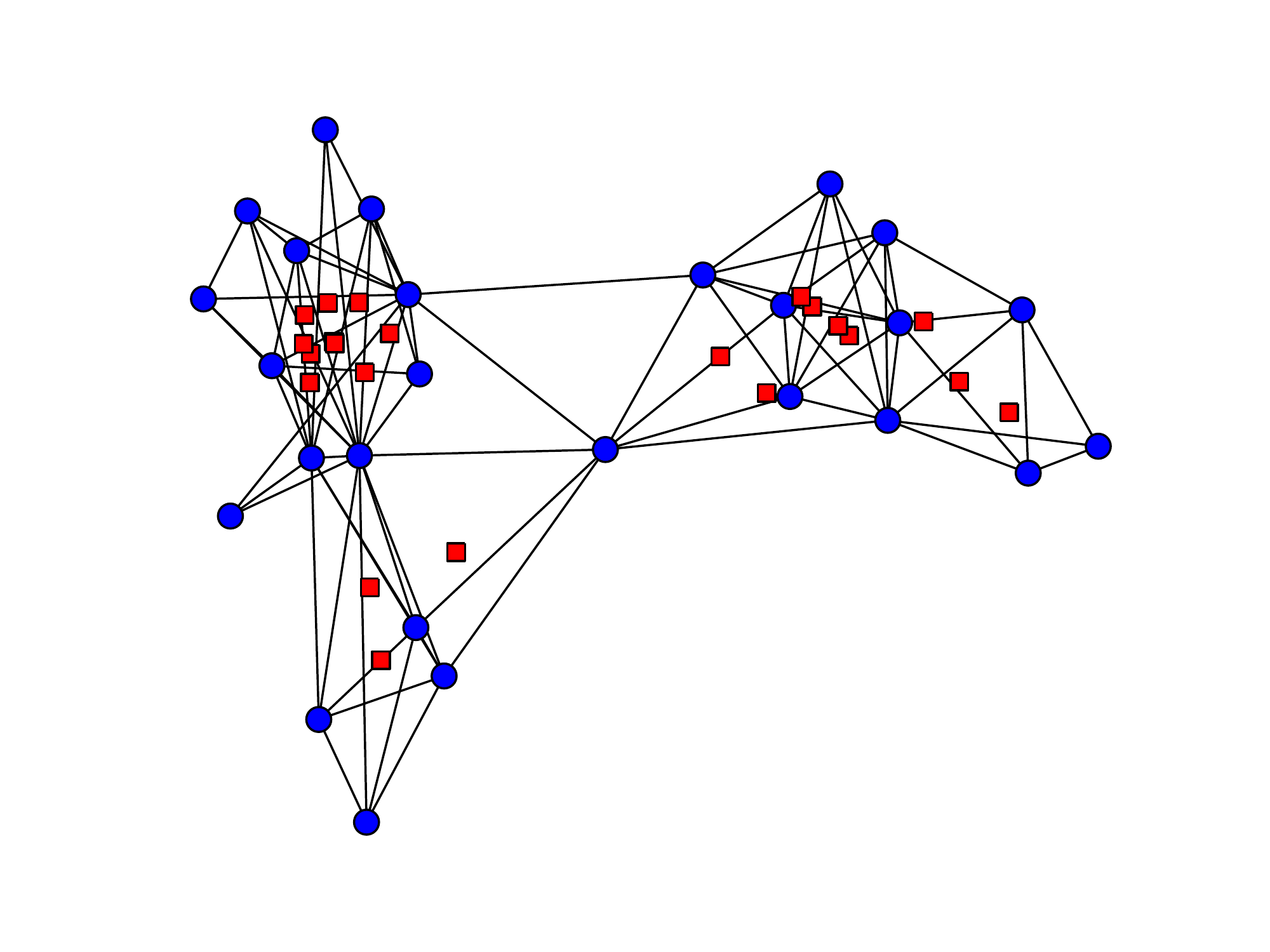}
        \label{2}
    }
    \subfigure[The corresponding clique graph of maximal cliques overlapping by at least 3 nodes.]{
        \includegraphics[width=70mm]{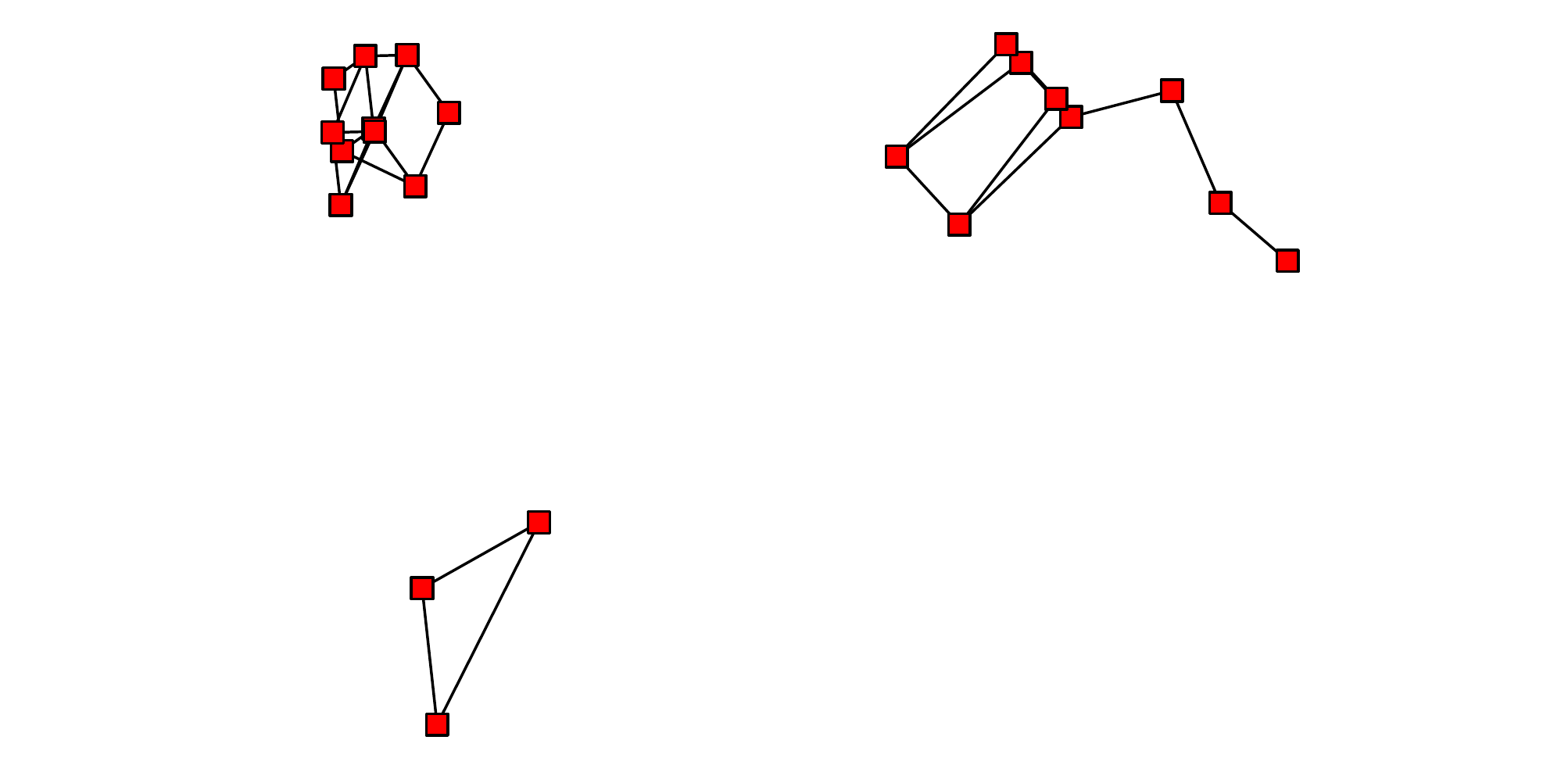}
        \label{3}
    }
    \subfigure[A minimal spanning forest of the clique graph.]{
        \includegraphics[width=70mm]{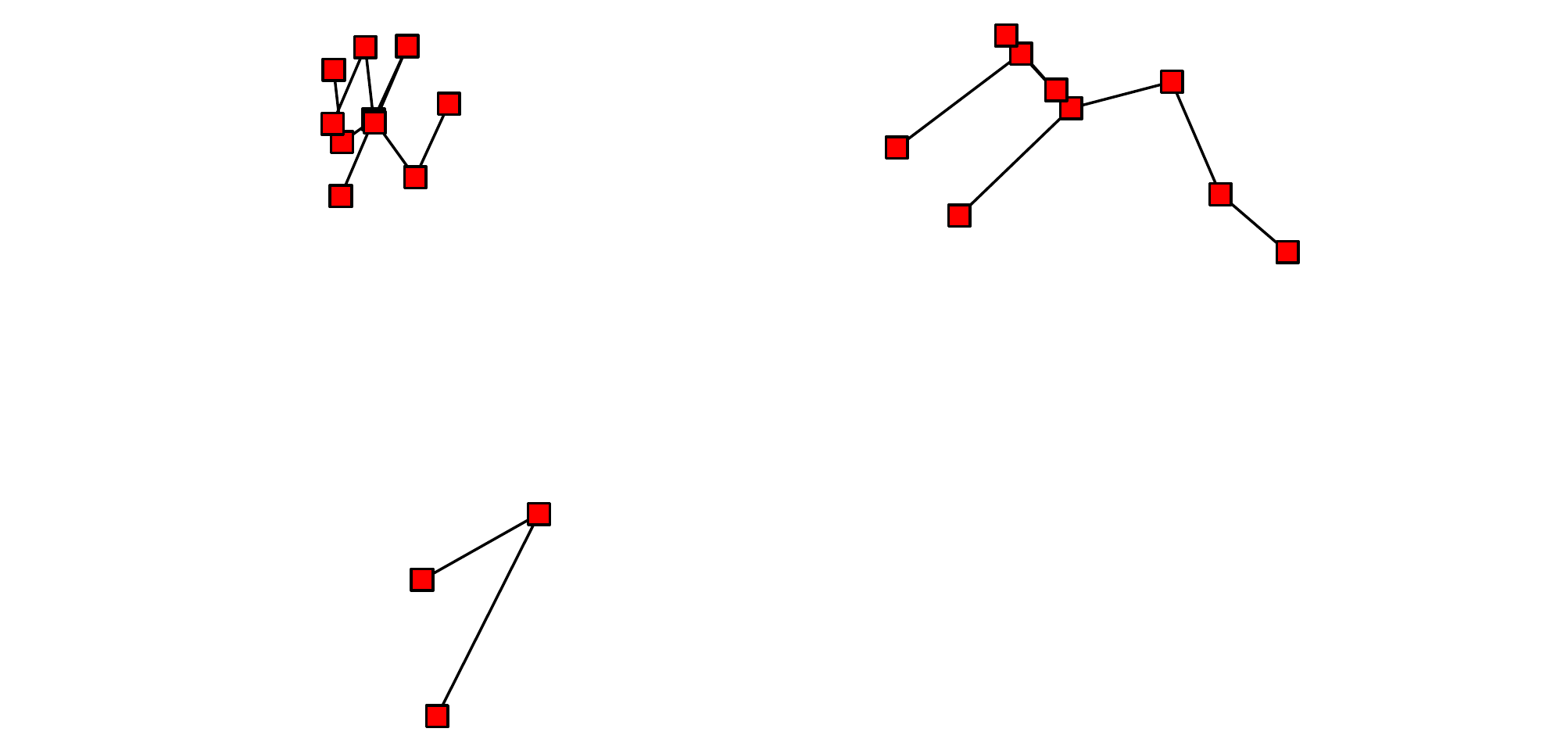}
        \label{4}
    }
\caption{A clique graph visualisation, showing the construction of the clique graph from a source network; the connected components in this graph, which correspond to clique percolations; and minimal spanning trees of these components -- the minimum set of edges in the clique graph which must be found, and which put an absolute lower bound the number of clique-clique intersection tests to be done in these methods.}
\label{minimalCliqueVisualisation}
\end{center}
\vspace{-10px}
\end{figure*}

The key point is that we do not first generate the full clique graph; instead we integrate the process of generating clique graph edges, with that of finding connected components; and only generate clique graph edges as we need them for the connected components calculation.
We can do this because, to calculate the $k$-clique percolations, we only need the minimal spanning forest of the clique graph (Fig \ref{minimalCliqueVisualisation}); any other edges in the clique graph are essentially wasted intersection tests.

We say that all the cliques are initially \emph{unvisited}, and
that they each become \emph{visited} as they are assigned to their component.
As the component expands, there is a set of cliques described as
the \emph{frontier}. The expansion proceeds until the frontier
is empty.
At each iteration, a clique is selected from the frontier (the `current frontier clique') and
its \emph{unvisited} neighbours in the clique graph are identified.
A `neighbouring' clique is a clique which has an intersection of at least $k$-1 nodes
with the current frontier clique; this corresponds to an edge in the clique graph.
These neighbours are moved into the frontier and marked as \emph{visited}
and our current frontier clique is moved out of the frontier, as it has now
been fully processed.
Eventually, there will be no unvisited neighbours of any of the frontier cliques
and the frontier will gradually become empty -- this completes the
identification of this component.
If we know that clique A is connected to clique B in the clique graph, and B to C,
then there is no need to test if A is connected to C; thus we can ignore cliques that have already been visited,
when searching for the neighbours of the current frontier clique.

Our two algorithms differ in how they identify the unvisited neighbouring cliques
of the current frontier node. Otherwise, they both use the framework which
has just been described.

One naive way to identify the neighbouring unvisited cliques of the current frontier clique is to
iterate through all the other cliques, testing the size of their intersection with the current frontier clique.
If the intersection has at least $k$-1 nodes \emph{and} if the clique is unvisited then
it is a neighbouring unvisited clique.
Our two algorithms embody different strategies to speed up this identification.

\subsection{Algorithm 1}
\label{SECalgorithm1}
\subsubsection{Node-to-cliques maps}
For a given current frontier clique, testing against every other clique is expensive.
We can vastly improve this, by realizing that two cliques can only neighbour each other in the clique graph, if they share at least one node (i.e.\ are adjacent by at least one node).
We maintain a list, for each node, of all the cliques that the node is present in.
Then, for each node in the current clique, we check the list of other cliques that node is in, to generate the set of cliques that overlap with the current clique by at least one node.


\subsubsection{Visiting the cliques}
The key step in Algorithm 1 is that as cliques are added to the current connected component, they are deleted from the map of nodes-to-cliques.
This is how a clique is marked as `visited', and not considered further. Pseudo-code is given in Listing \ref{python}.

\definecolor{dkgreen}{rgb}{0,0.6,0}
\definecolor{gray}{rgb}{0.5,0.5,0.5}
\definecolor{mauve}{rgb}{0.58,0,0.82}

\lstset{ %
  language=Python,                
  basicstyle=\footnotesize,           
  numbers=left,                   
  numberstyle=\tiny\color{gray},  
  stepnumber=2,                   
  numbersep=5pt,                  
  backgroundcolor=\color{white},      
  showspaces=false,               
  showstringspaces=false,         
  showtabs=false,                 
  frame=single,                   
  rulecolor=\color{black},        
  tabsize=2,                      
  captionpos=b,                   
  breaklines=true,                
  breakatwhitespace=false,        
  title=\lstname,                   
  keywordstyle=\color{blue},          
  commentstyle=\color{dkgreen},       
  stringstyle=\color{mauve},         
  escapeinside={\%*}{*)},            
  morekeywords={*,...}               
}

\begin{lstlisting}[label={python},float=*,caption={Python-like pseudo-code description of simple algorithm for improved clique percolation.}]
    for clique in cliques:
        if not clique in cliques_to_components_dict:
            current_component += 1
            cliques_to_components_dict[clique] = current_component
            frontier = set()
            frontier.add(clique)

            while len(frontier) > 0:
                current_clique = frontier.pop()
                for neighbour in get_unvisited_adjacent_cliques(current_clique, nodes_to_cliques_dict):
                    if len(current_clique.intersection(neighbour)) >= (k-1):
                        cliques_to_components_dict[neighbour] = current_component
                        frontier.add(neighbour)
                        for node in neighbour:
                            nodes_to_cliques_dict[node].remove(neighbour)
\end{lstlisting}

\subsection{Experiments}
\ifthenelse {\boolean{longVersion}} {
We performed several benchmarking experiments, of a C++ version of our algorithm, against the binary implementations of both SCP and CFinder v2.0.5.
We do not show benchmark results for CFinder, as, similar to Kumpula et al. \cite{kumpula2008sequential} we found that it was generally very much slower than SCP.
As such, all benchmarks will be shown against SCP.
}

\subsubsection{GN benchmarks}
\begin{figure}[!t]
\begin{center}
    \subfigure[Clique size distribution for\newline Simmons81 Facebook network.]{
        \includegraphics[width=41mm]{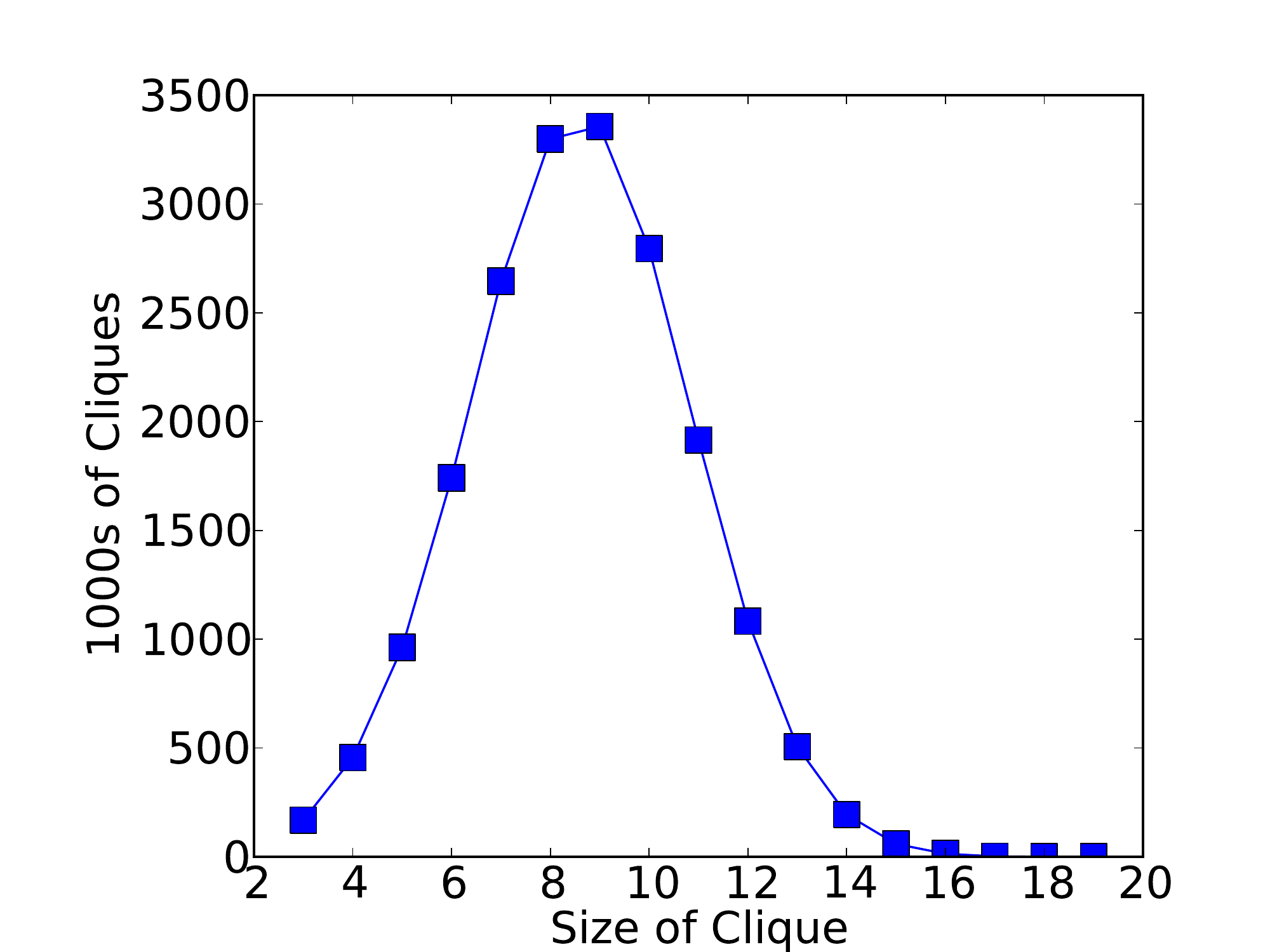}
        \label{C1}
    }
    \subfigure[Maximal clique size distribution for Simmons81 Facebook network.]{
        \includegraphics[width=41mm]{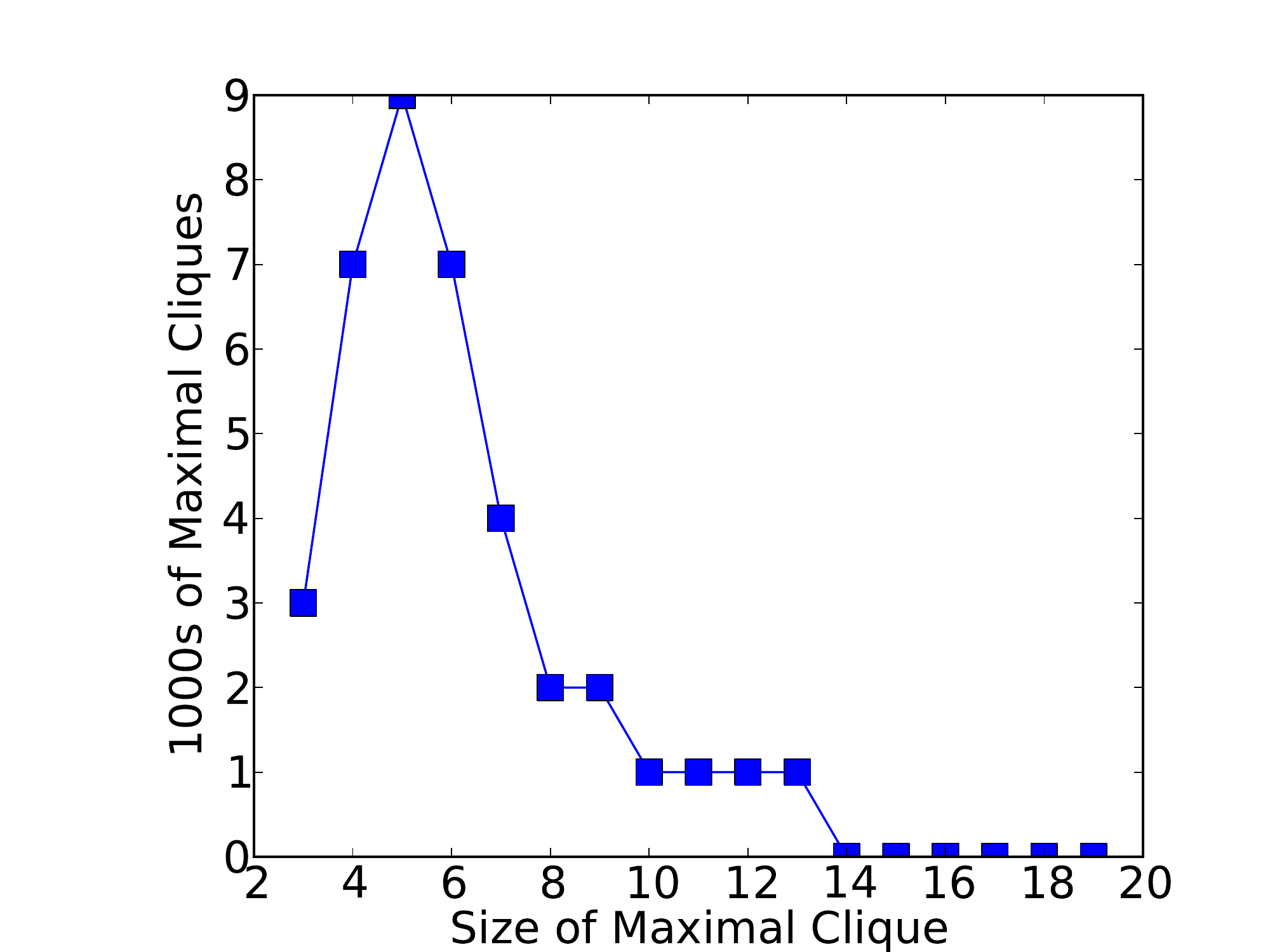}
        \label{C2}
    }
    \subfigure[Clique size distribution,\newline GN benchmark.]{
        \includegraphics[width=41mm]{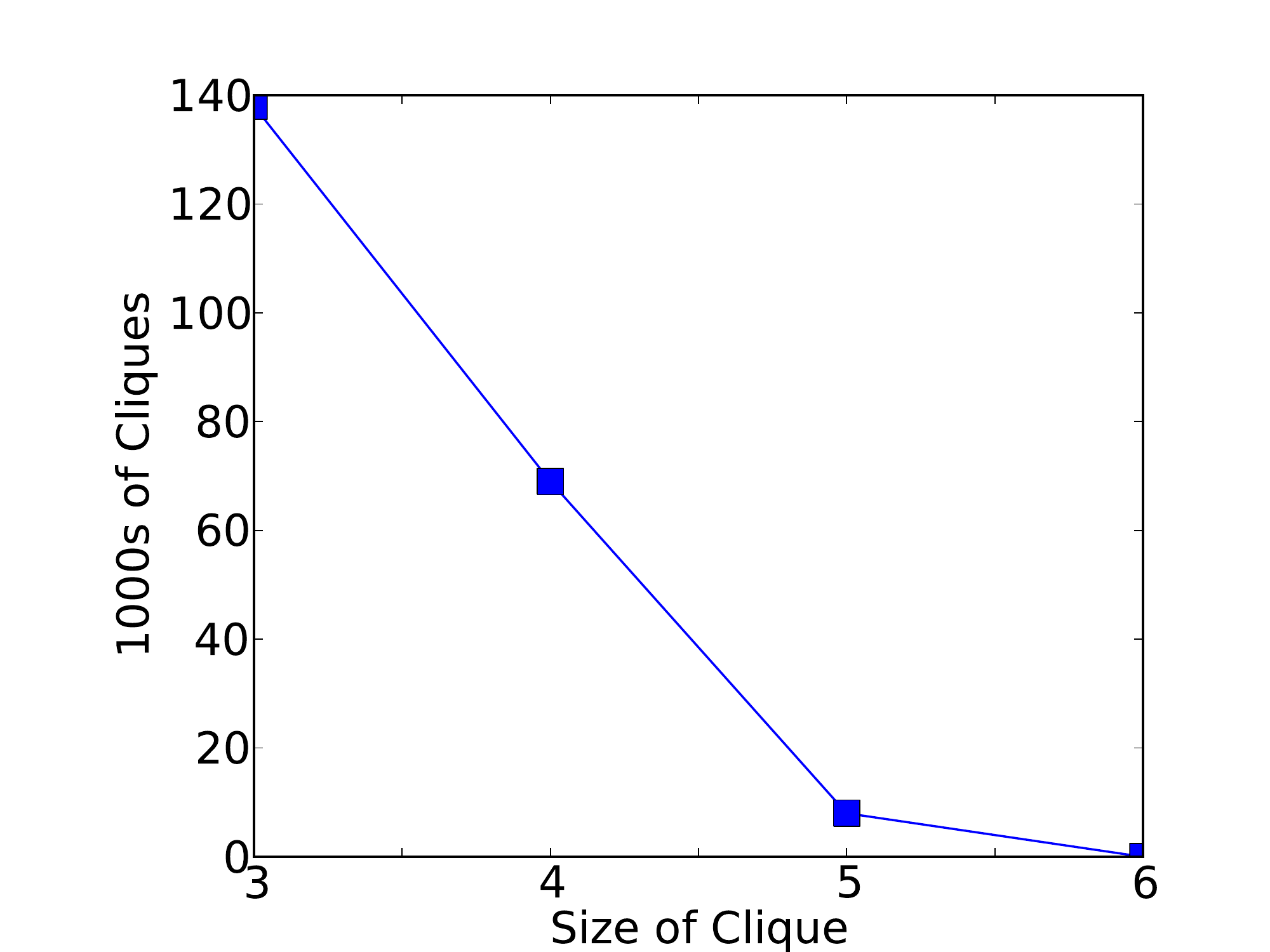}
        \label{C3}
    }
    \subfigure[Maximal clique size distribution, GN benchmark.]{
        \includegraphics[width=41mm]{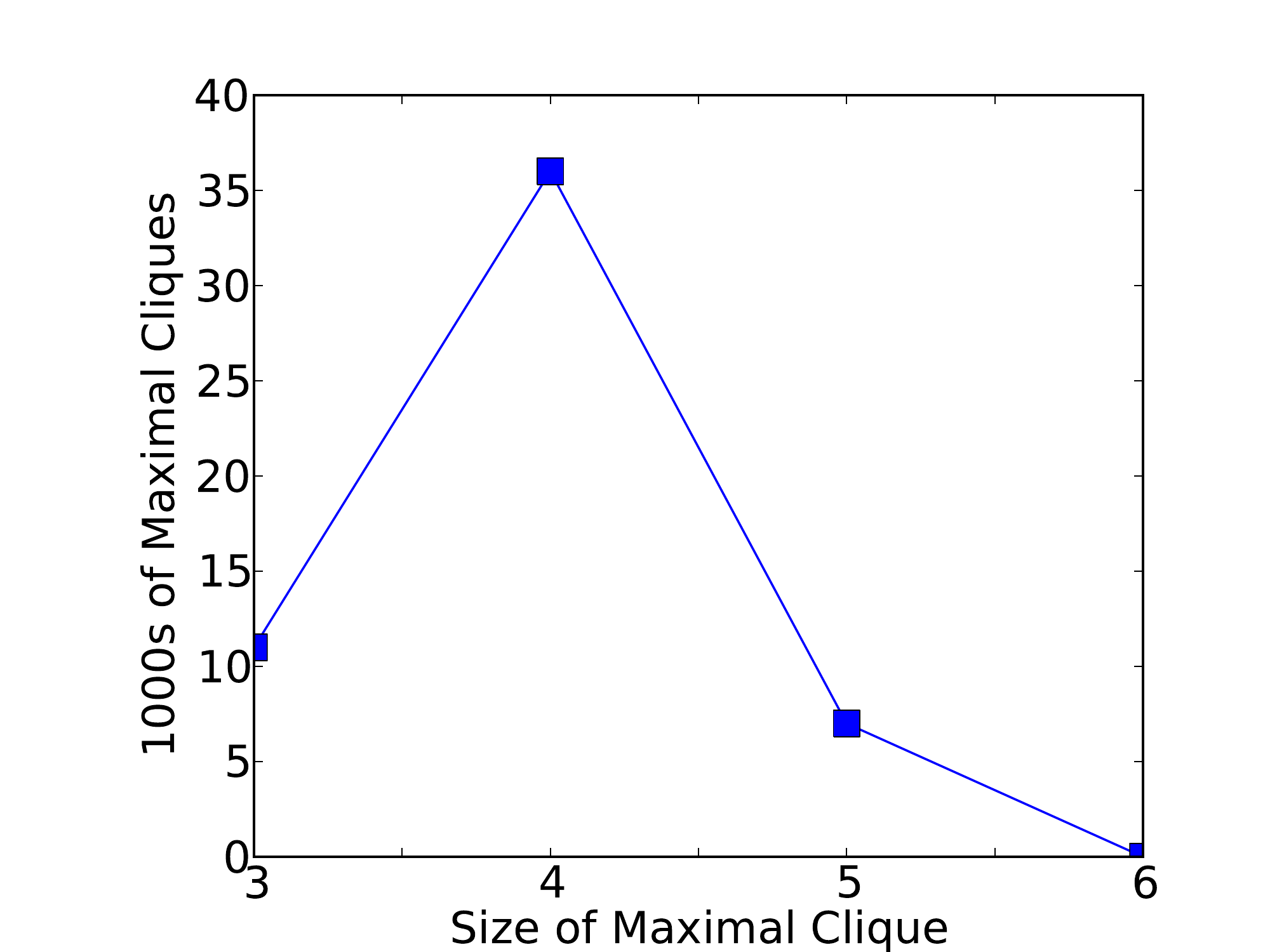}
        \label{C4}
    }
\caption{Clique and maximal clique size distributions, for benchmark and empirical networks. The benchmark network is a GN network, of 10,000 nodes, constructed to a similar specification as in \cite{kumpula2008sequential}. The Facebook network is that of the Simmons81 Facebook 100 dataset, and is typical of Facebook 100 data. We note that there are typically very many more cliques than maximal cliques; that the peak of the distributions for the empirical network is further to the left for the maximal cliques, than for all cliques; and that more cliques of much larger sizes are found in empirical data, than in these GN benchmarks, despite this particular GN benchmark network's larger size. }
\label{CliqueSizeDistributions}
\end{center}
\vspace{-12px}
\end{figure}

After Kumpula et al. \cite{kumpula2008sequential}, we performed benchmarks based on the Girvan-Newman synthetic benchmark, creating networks of increasing size, containing increasing numbers of communities, each of 32 nodes.
We first note that these synthetic benchmarks are poor proxies for many real world social networks, in terms of the clique size distributions contained within them. 
Figure \ref{CliqueSizeDistributions} shows a typical example of this difference in the size distributions of cliques, and maximal cliques, in these GN networks, versus the Facebook networks; the Facebook networks have more and larger cliques and maximal cliques.
While the main advantage of our method over SCP is on networks with larger numbers of cliques, which these GN networks do not contain, we present, in Figure \ref{GNResults}, present performance results as we increase the number of nodes in the GN network.
At $k=4$ our method, while slightly slower than SCP, is competitive, and scales similarly.
However, when we start to look at higher values of $k$, we notice that the scaling behaviour of SCP degrades considerably faster than our method.

\ifthenelse {\boolean{longVersion}} {

\begin{figure*}[!htb]
\begin{center}
    \subfigure[Runtimes for K=4.]{
        \includegraphics[width=60mm]{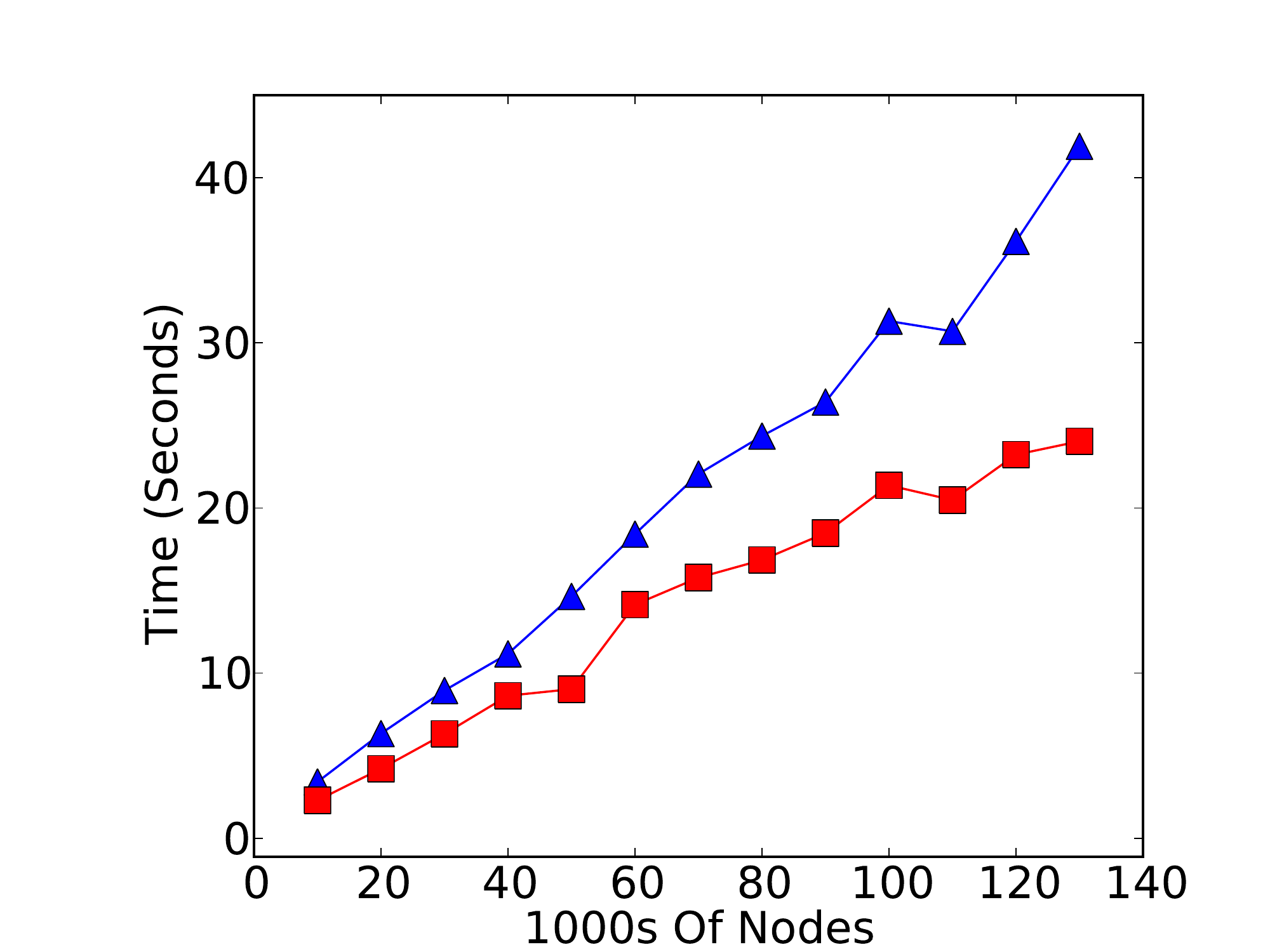}
        \label{C1Times}
    }
    \subfigure[Runtimes for K=5.]{
        \includegraphics[width=60mm]{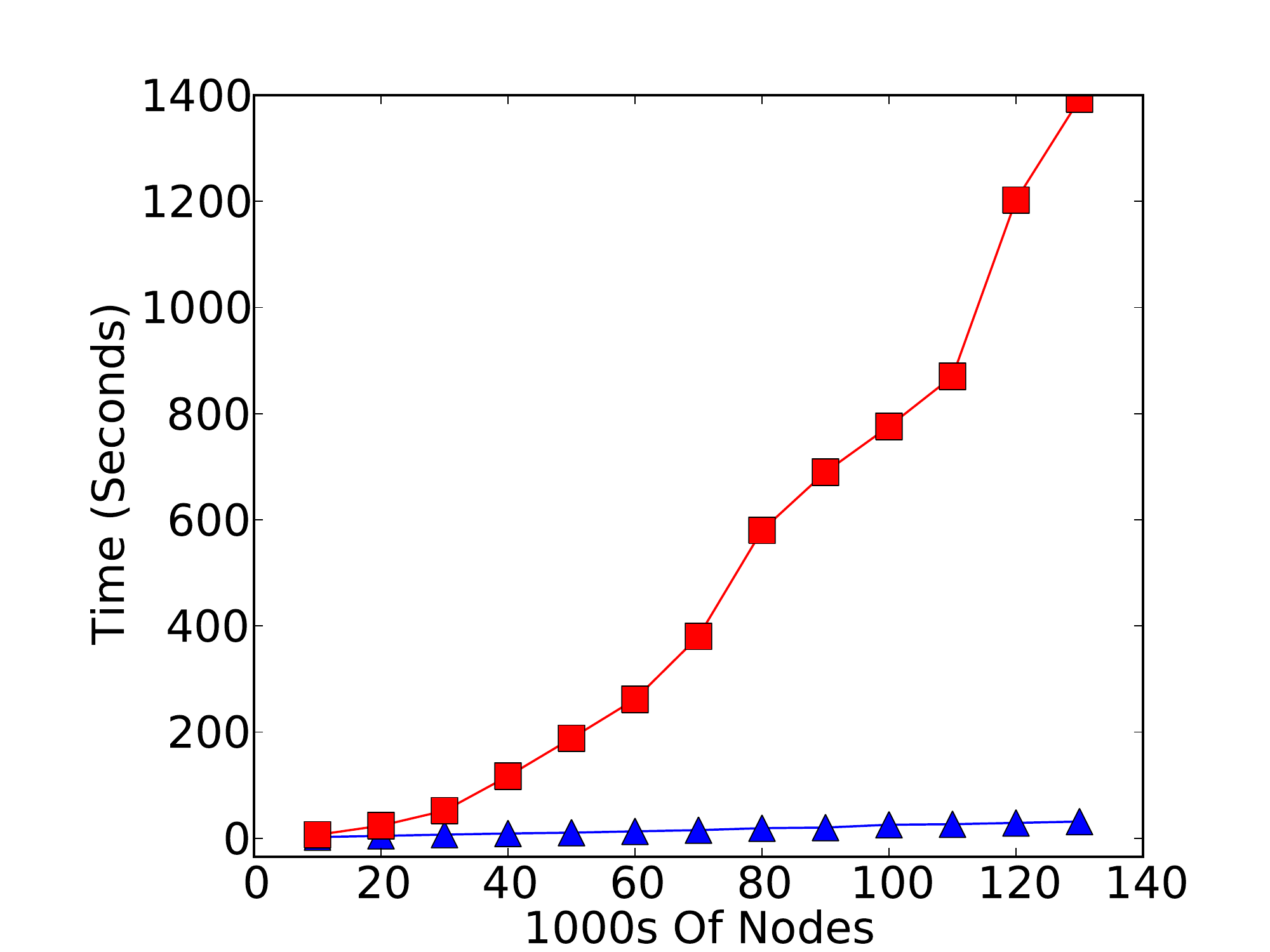}
        \label{C2Times}
    }
\caption{GN benchmark results, on a similar benchmarking setup to \cite{kumpula2008sequential}. SCP runtimes are shown by red squares; runtimes of our method are shown by blue triangles. Runtimes are observed to be similar when $k=4$, with SCP marginally ahead; but for $k=5$ our method appears to do much better. There are insufficient numbers of cliques of size 6 or greater present to allow higher values of $k$ to be examined on the GN benchmarks.}
\label{GNResults}
\end{center}
\vspace{-12px}
\end{figure*}

}
{
\begin{figure}[!htb]
\begin{center}
    \subfigure[Runtimes for K=4.]{
        \includegraphics[width=41mm]{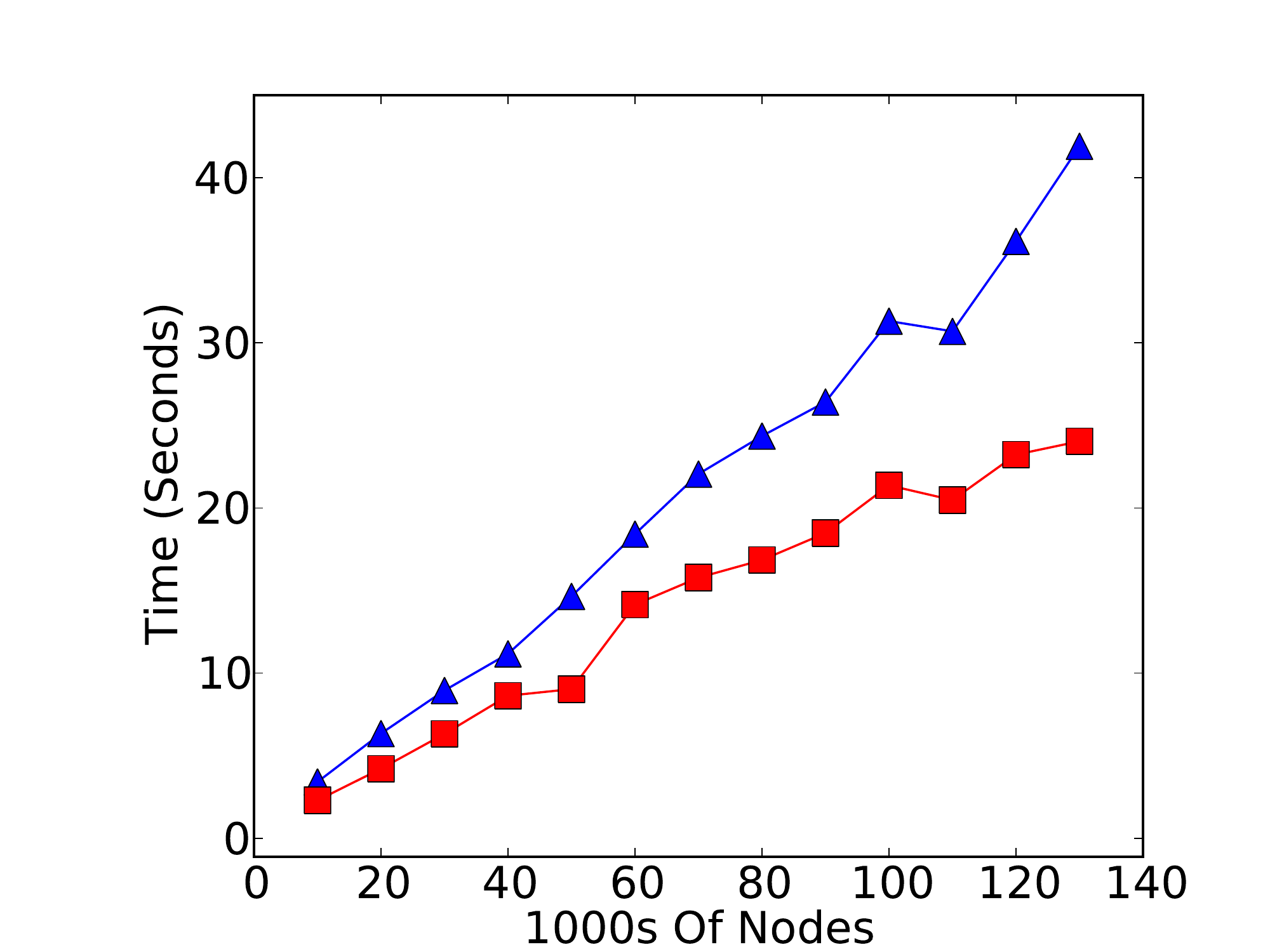}
        \label{C1Times}
    }
    \subfigure[Runtimes for K=5.]{
        \includegraphics[width=41mm]{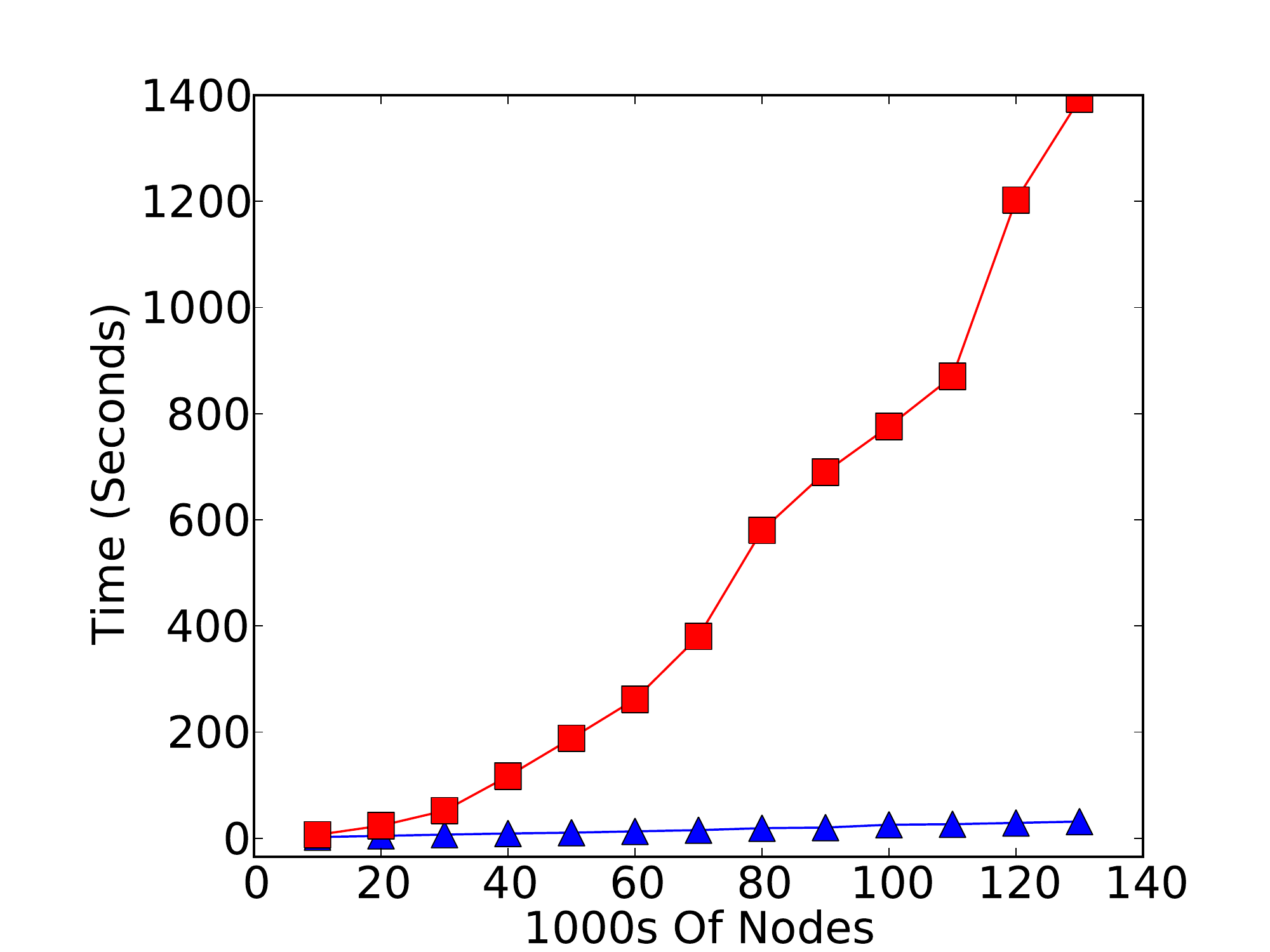}
        \label{C2Times}
    }
\caption{GN benchmark results, on a similar benchmarking setup to \cite{kumpula2008sequential}. SCP runtimes are shown by red squares; runtimes of our method are shown by blue triangles. Runtimes are observed to be similar when $k=4$, with SCP marginally ahead; but for $k=5$ our method appears to do much better. There are insufficient numbers of cliques of size 6 or greater present to allow higher values of $k$ to be examined on the GN benchmarks.}
\label{GNResults}
\end{center}
\vspace{-12px}
\end{figure}
}

\subsubsection{CondMat benchmark}
An example of using $k$-clique percolation to find communities in co-authorship data in the field of condensed matter physics, is described by Palla et al. \cite{palla2005uncovering}.
The SCP paper \cite{kumpula2008sequential} extends this benchmark, comparing the times of SCP and CFinder for finding $k$-clique percolations, in the Arxiv Condensed Matter e-print archive.
We produce a similar benchmark, using the Arxiv ca-CondMat network from the SNAP dataset collection~\cite{SNAP}.
Figure \ref{CondMat} shows the runtime of SCP against that of our implementation, on this network, for varying values of $k$.
Note that $y$-axis, showing runtime, is logarithmic.
Both implementations quickly complete for low values of $k$, and, as shown in \cite{kumpula2008sequential}, far outperform CFinder; but as the value of $k$ increases, the runtime of SCP deteriorates.
As noted in the original SCP work ``for networks containing large cliques the SCP method performs best for rather small values of k'' \cite{kumpula2008sequential}.
Our method, by contrast, seems well suited to this particular type of network, and performs quickly, even as $k$ is increased.

\ifthenelse {\boolean{longVersion}} {

\begin{figure}[!htb]
\begin{center}
        \includegraphics[width=60mm]{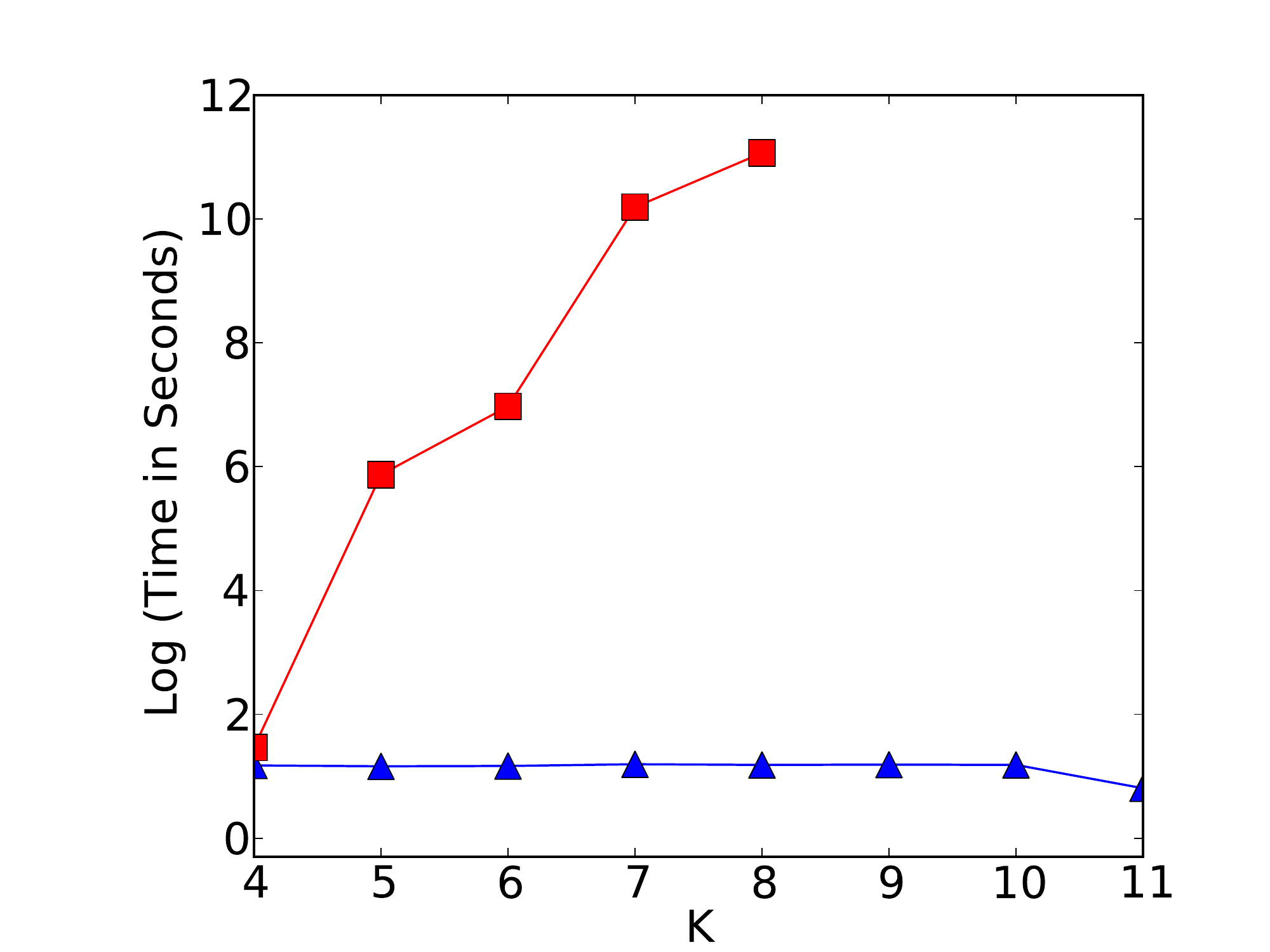}
\caption{Results on the Arxiv CondMat network, from the SNAP dataset collection \cite{SNAP}.  This network was also benchmarked in the work of Kumpula et al. \cite{kumpula2008sequential}.  This is a network which, perhaps due to the fact it is a one-mode projection of a bipartite author network, our algorithm seems to be particularly well on, as K grows, in contrast to SCP. The SCP implementation was terminated, on values of K larger than 8, for exceeding available memory. Any single process exceeding 75GB of RAM was terminated.}
\label{CondMat}
\end{center}
\end{figure}

\begin{figure}[!htb]
\begin{center}
        \includegraphics[width=60mm]{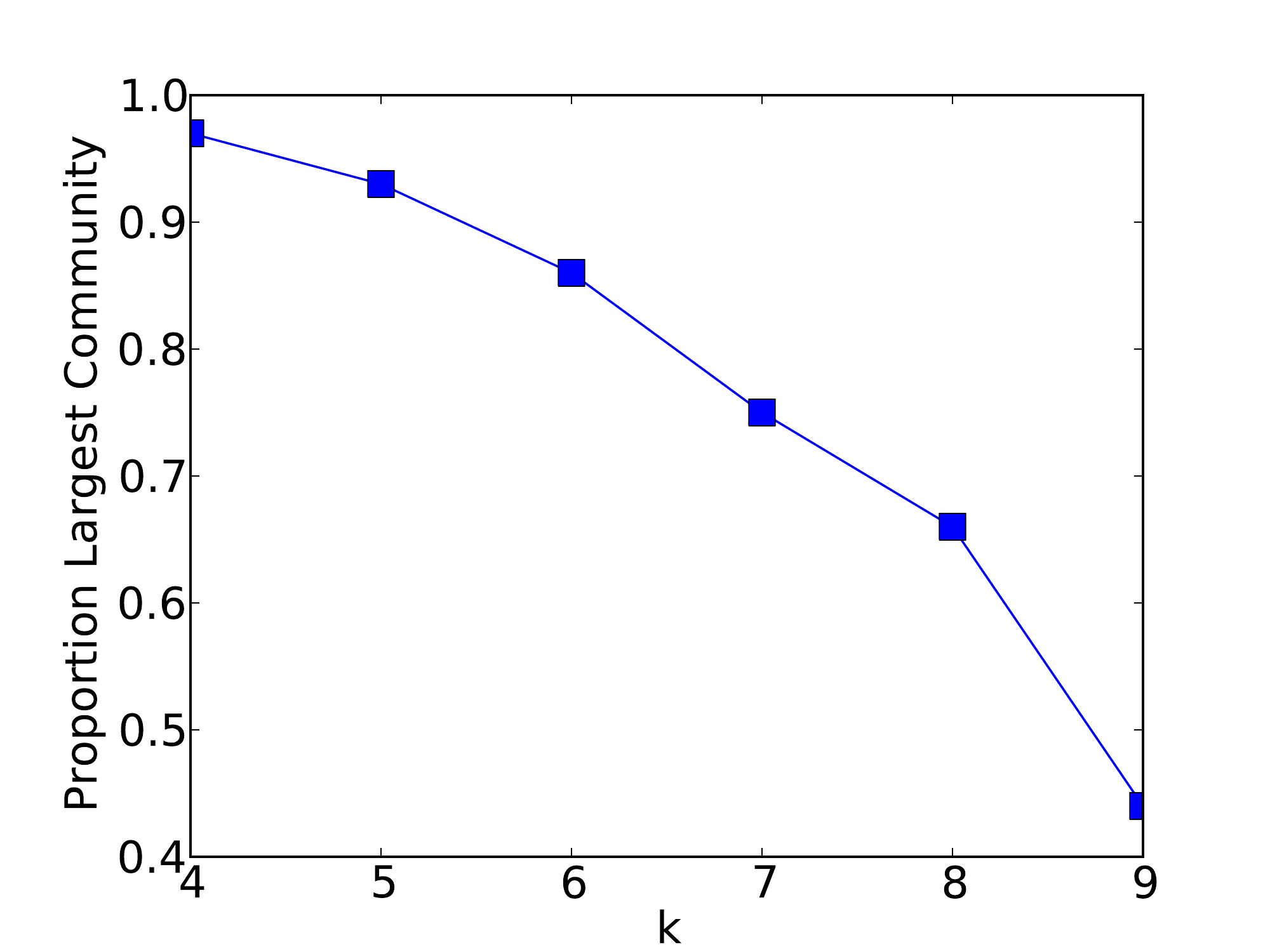}
\caption{The proportion of nodes assigned to any community, that are assigned to the largest community, for each value of $k$, on the Mich67 Facebook network. Clearly, on this particular network, low values of $k$ simply find a core structure of the network with pervasive overlap, and do not serve to reveal modular units; larger values of $k$ are required for $k$-clique percolation to be in any way meaningful on these networks.}
\label{MichProp}
\end{center}
\end{figure}

}
{
\begin{figure}[!htb]
\begin{center}
    \subfigure[]{
        \includegraphics[width=41mm]{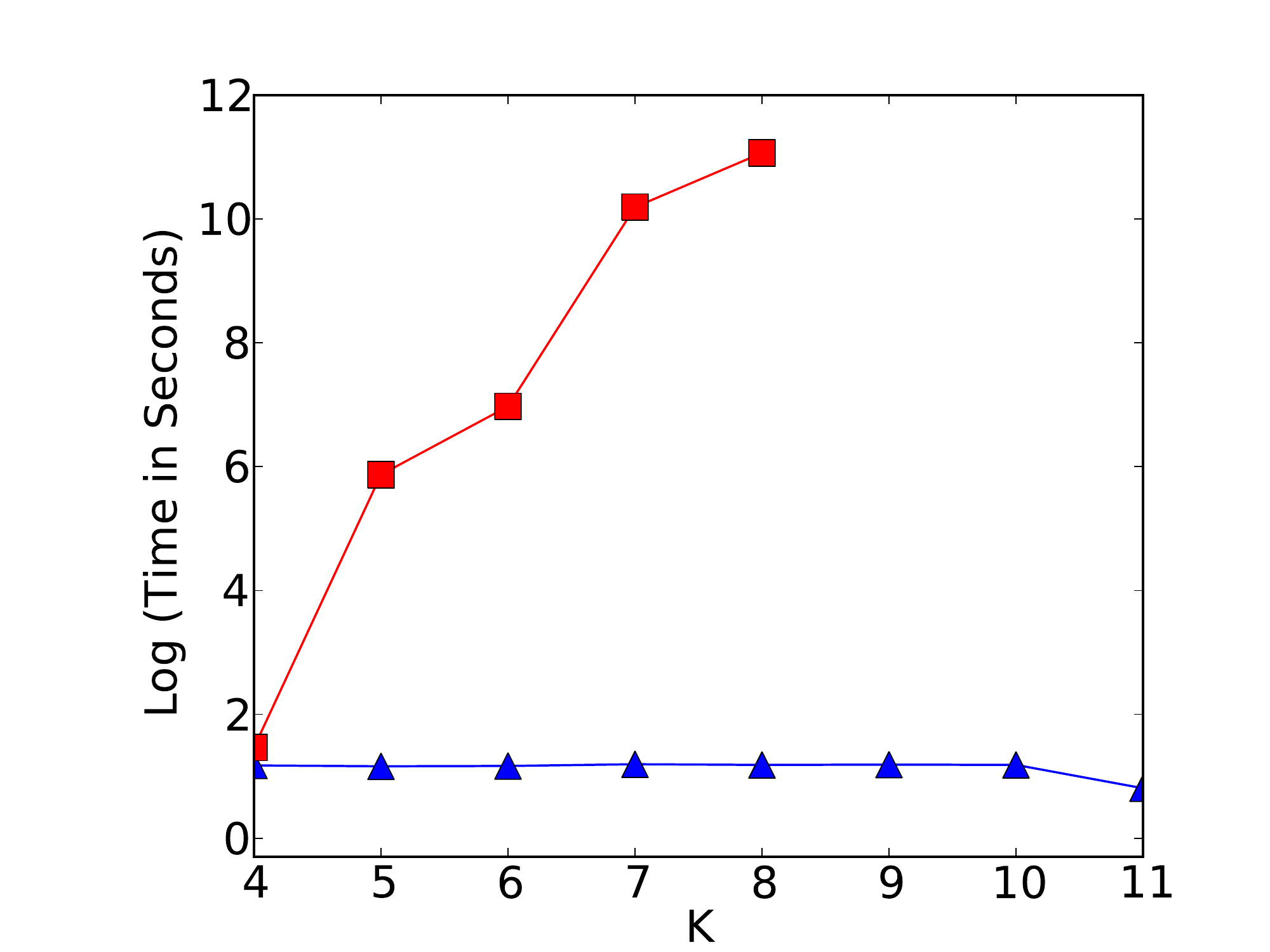}
        \label{CondMat}
    }
    \subfigure[]{
        \includegraphics[width=41mm]{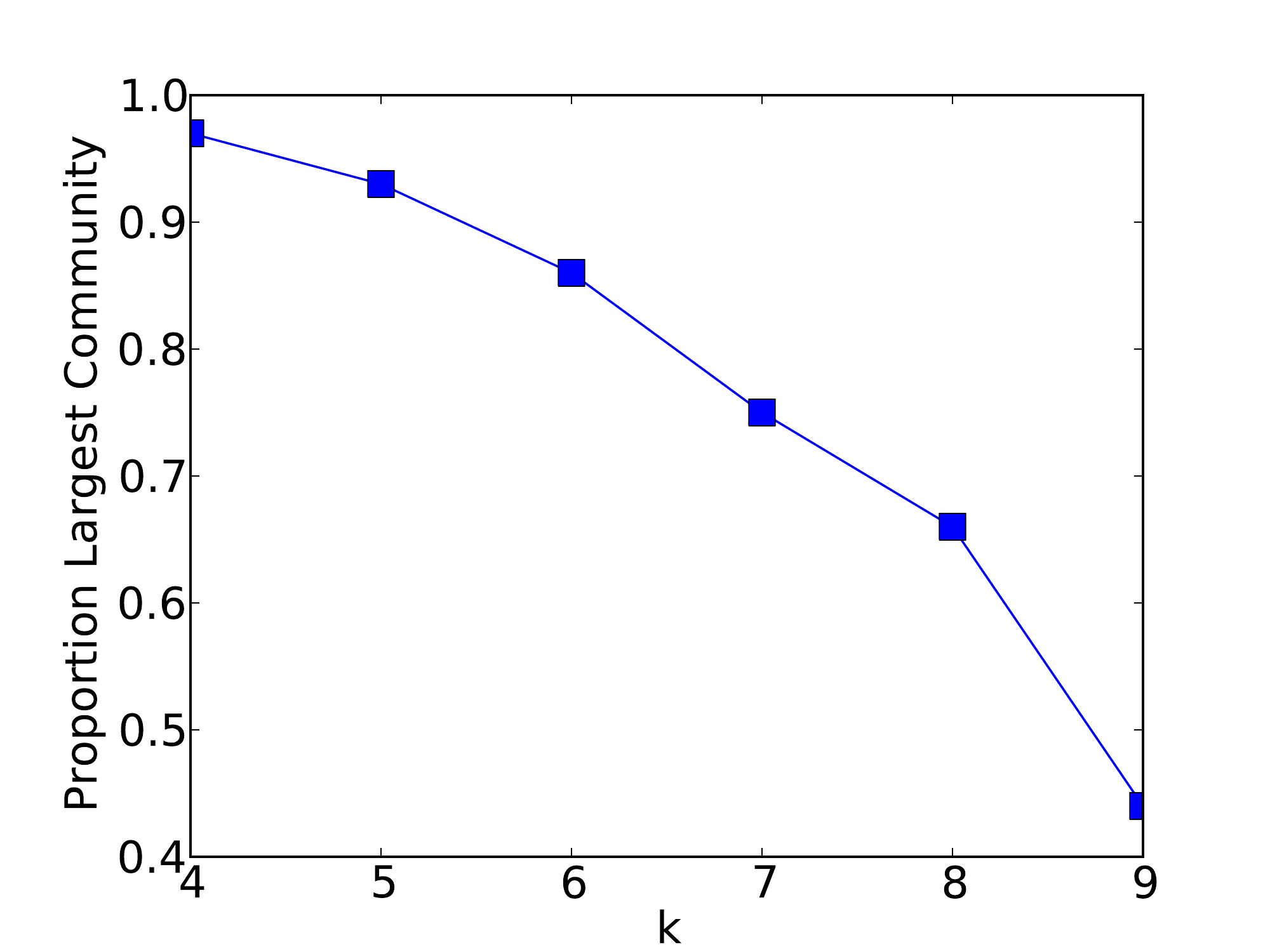}
        \label{MichProp}
    }
\caption{(a) Results on the Arxiv CondMat network, from the SNAP dataset collection \cite{SNAP}.  This network was also benchmarked in the work of Kumpula et al. \cite{kumpula2008sequential}.  This is a network which, perhaps due to the fact it is a one-mode projection of a bipartite author network, our algorithm seems to be particularly well on, as K grows, in contrast to SCP. The SCP implementation was terminated, on values of K larger than 8, for exceeding available memory. Any single process exceeding 75GB of RAM was terminated. \newline(b) The proportion of nodes assigned to any community, that are assigned to the largest community, for each value of $k$, on the Mich67 Facebook network. Clearly, on this particular network, low values of $k$ simply find a core structure of the network with pervasive overlap, and do not serve to reveal modular units; larger values of $k$ are required for $k$-clique percolation to be in any way meaningful on these networks.}
\end{center}
\vspace{-12px}
\end{figure}

}

\subsubsection{Empirical social network data}
We now produce a set of empirical benchmarks, not previously investigated in the $k$-clique percolation literature.
We analyse the performance of SCP against our algorithm, for various values of $k$, on the ten smallest of the Facebook 100 networks.
This allows us easily obtain results for higher values of $k$ where SCP cannot complete in reasonable lengths of time.
Figure \ref{FacebookBenchmarkResults} shows performance results for four values of $k$ across these 10 Facebook networks.
These empirical networks vary in their individual performance characteristics. 
In common with previous work on $k$-clique percolation, it is difficult to give meaningful closed form analysis of expected performance characteristics in terms of more general network parameters, such as the number of nodes, or edges; but the trend is clearly that as $k$ increases, the performance of our method versus SCP clearly increases, across all networks.
Even for the lowest value of $k$ that we examine, our method finishes within a few minutes of SCP, while as the value of $k$ increases, it becomes difficult to obtain runtimes for SCP.
Furthermore, in additional to increasing runtime, the memory usage of SCP's data structures, which allow for fast intersection tests, grows prohibitive. 

When analysing these networks, we often find that, for low values of $k$, a giant connected component emerges in the network, due to the pervasively overlapping structure of community in these networks. 
Figure \ref{MichProp} shows, for each value of $k$ on the Mich67 Facebook network,
the proportion of nodes
which are in the largest $k$-clique percolation community
with respect to the number of nodes which have been assigned to at least one $k$-clique percolation community.
As can be clearly seen from this chart, smaller values of $k$ do not usefully find modular structure in the network.
An investigation of how meaningful these structures are as communities is, however, beyond the scope of this computational work; but this question cannot even be investigated without an algorithm which runs reasonably quickly for higher values of $k$. 
\ifthenelse {\boolean{longVersion}} {
This particular chart should not be taken as characteristic of all pervasively overlapping networks; while we do commonly find that for low values of $k$, a giant percolation component is present in the Facebook 100 datasets, the nature of how this changes as $k$ increases, varies from network to network.
}

\ifthenelse {\boolean{longVersion}} {
\begin{figure*}[!htb]
\begin{center}
    \subfigure[K=4 (Note smaller y-axis)]{
        \includegraphics[width=60mm]{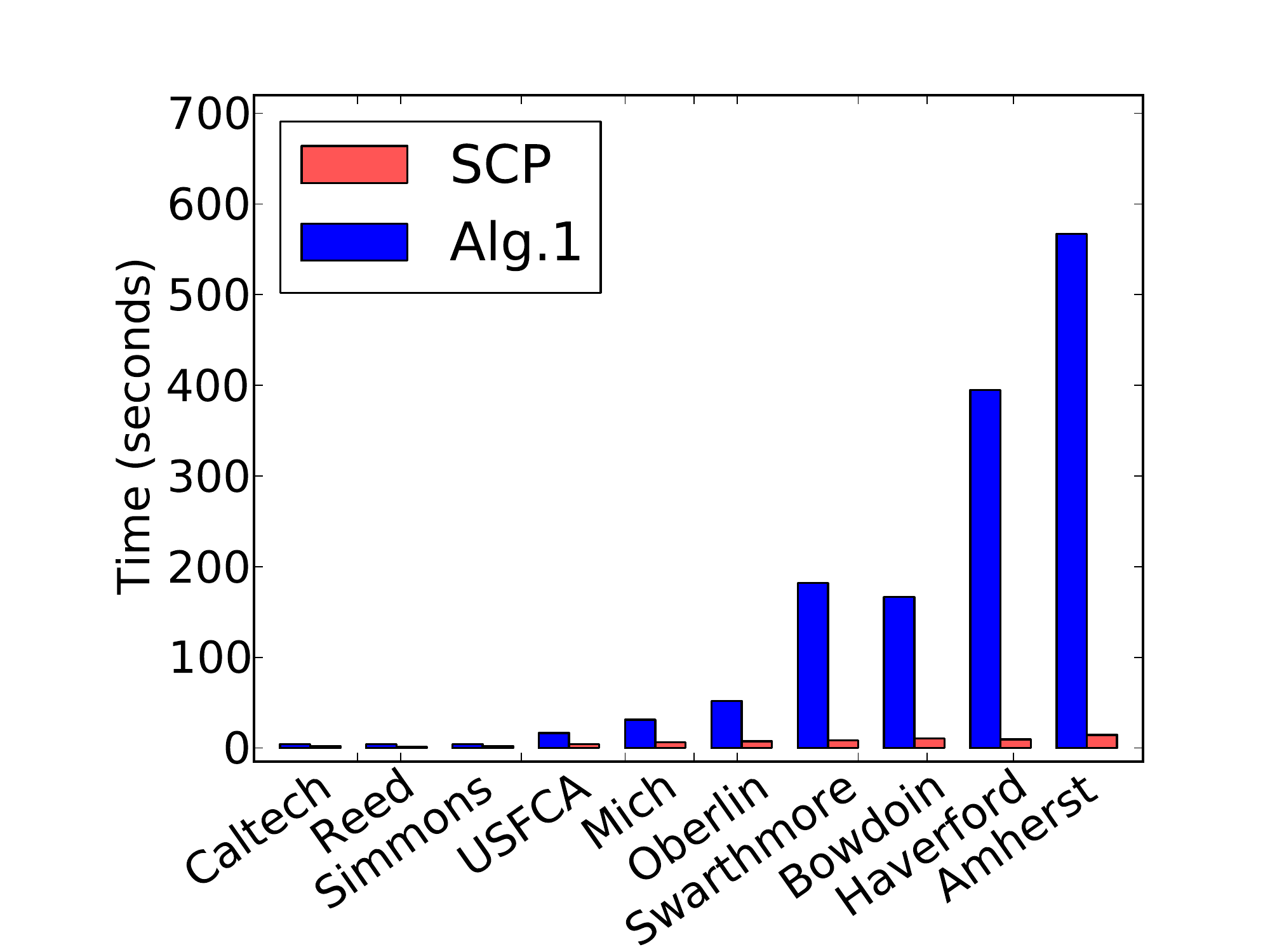}
        \label{K4}
    }
    \subfigure[K=5]{
        \includegraphics[width=60mm]{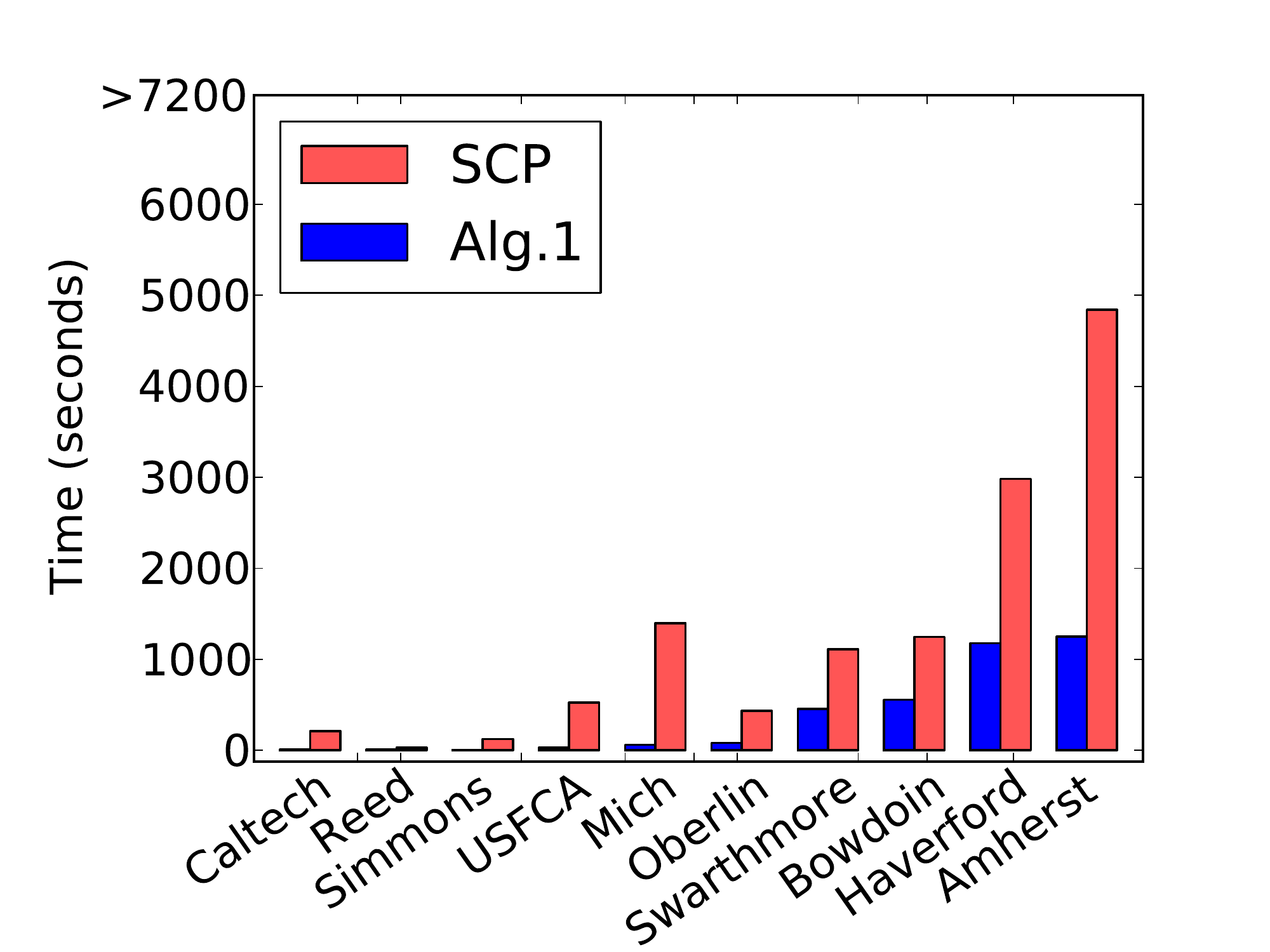}
        \label{K5}
    }
    \subfigure[K=6]{
        \includegraphics[width=60mm]{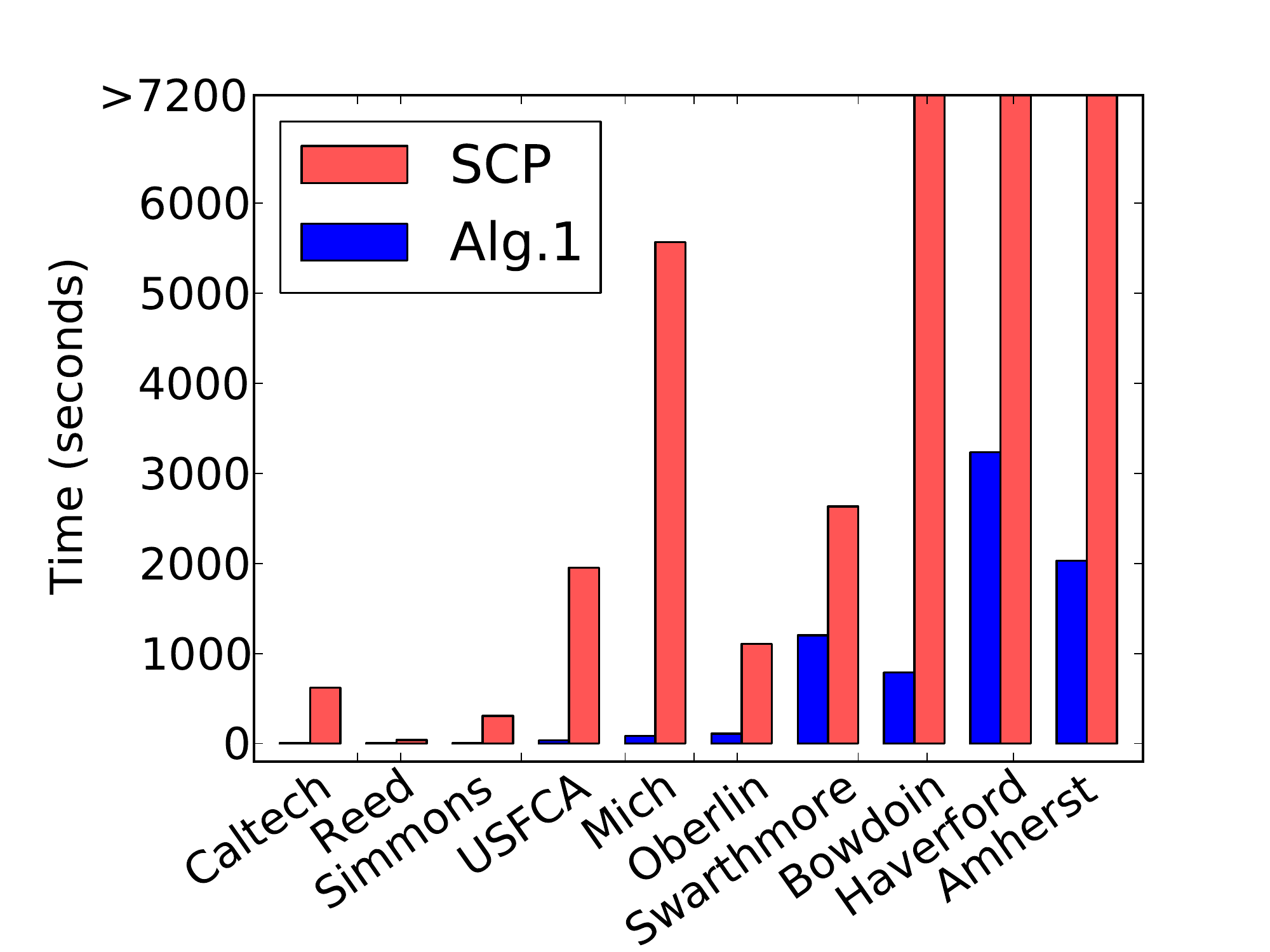}
        \label{K6}
    }
    \subfigure[K=7]{
        \includegraphics[width=60mm]{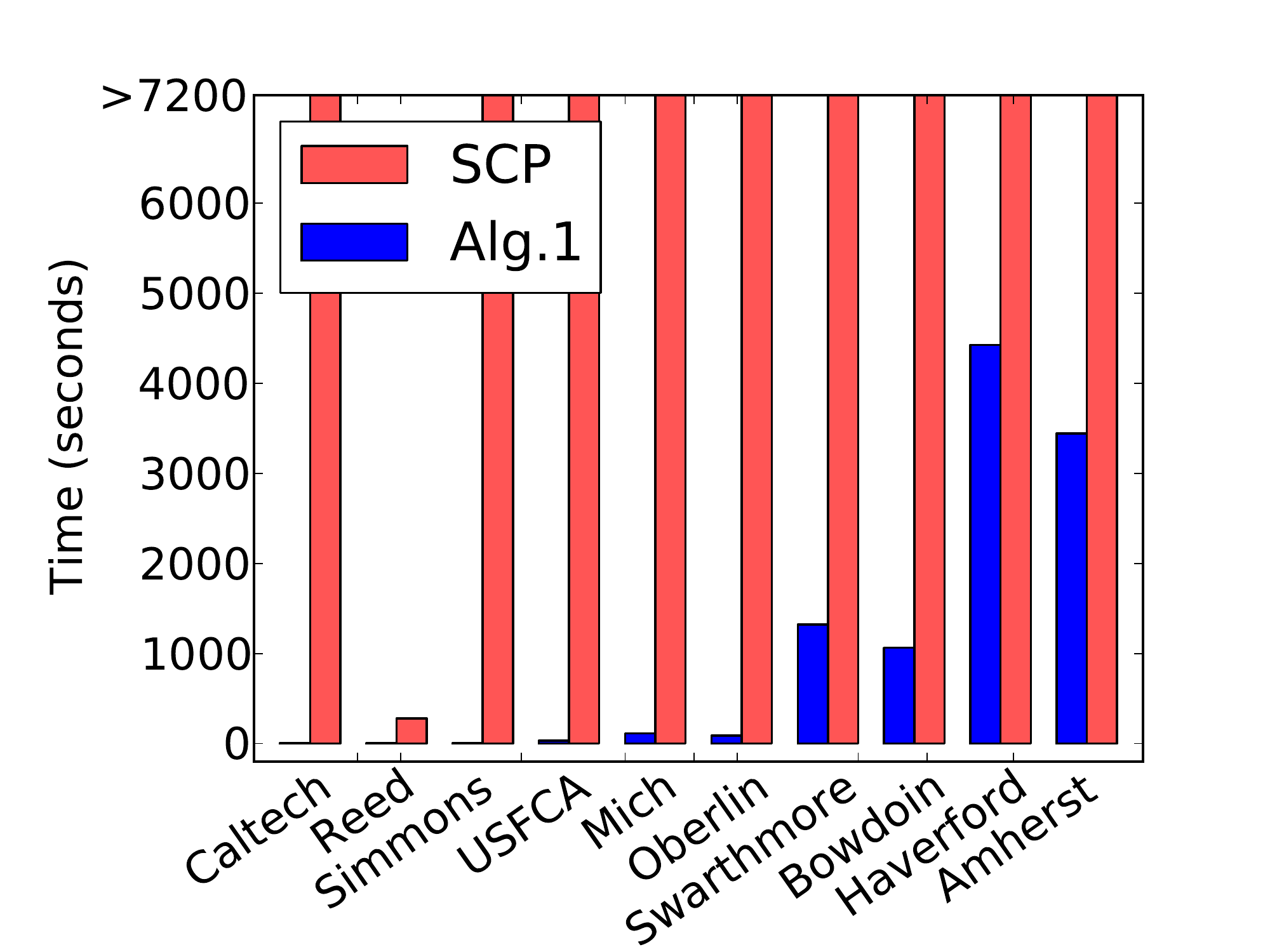}
        \label{K7}
    }
\caption{Benchmark results on the smallest 10 Facebook 100 graphs, for values of $k$ between $4$ and $7$. Red columns denote SCP runtimes, blue denote Algorithm 1. Processes were terminated if they failed to complete within 2 hours (7200 seconds). Running SCP for additional time sometimes resulted in it using more than 50GB of RAM; the SCP binary implementation consumes a vast amount of RAM for some networks and values of $k$, presumably for its disjoint-set forest fast intersection data structure. Our method never uses more than a small multiple of the amount used to store the maximal cliques; never more than ~512MB in these experiments. Experiments are run on a multiprocessor machine, in random sequence. SCP outperforms our algorithm for $k=4$, but performs substantially worse, in both time and space requirements, as values of $k$ get larger.}
\label{FacebookBenchmarkResults}
\end{center}
\vspace{-10px}
\end{figure*}

}
{
\begin{figure}[!htb]
\begin{center}
    \subfigure[K=4 (Note smaller y-axis)]{
        \includegraphics[width=41mm]{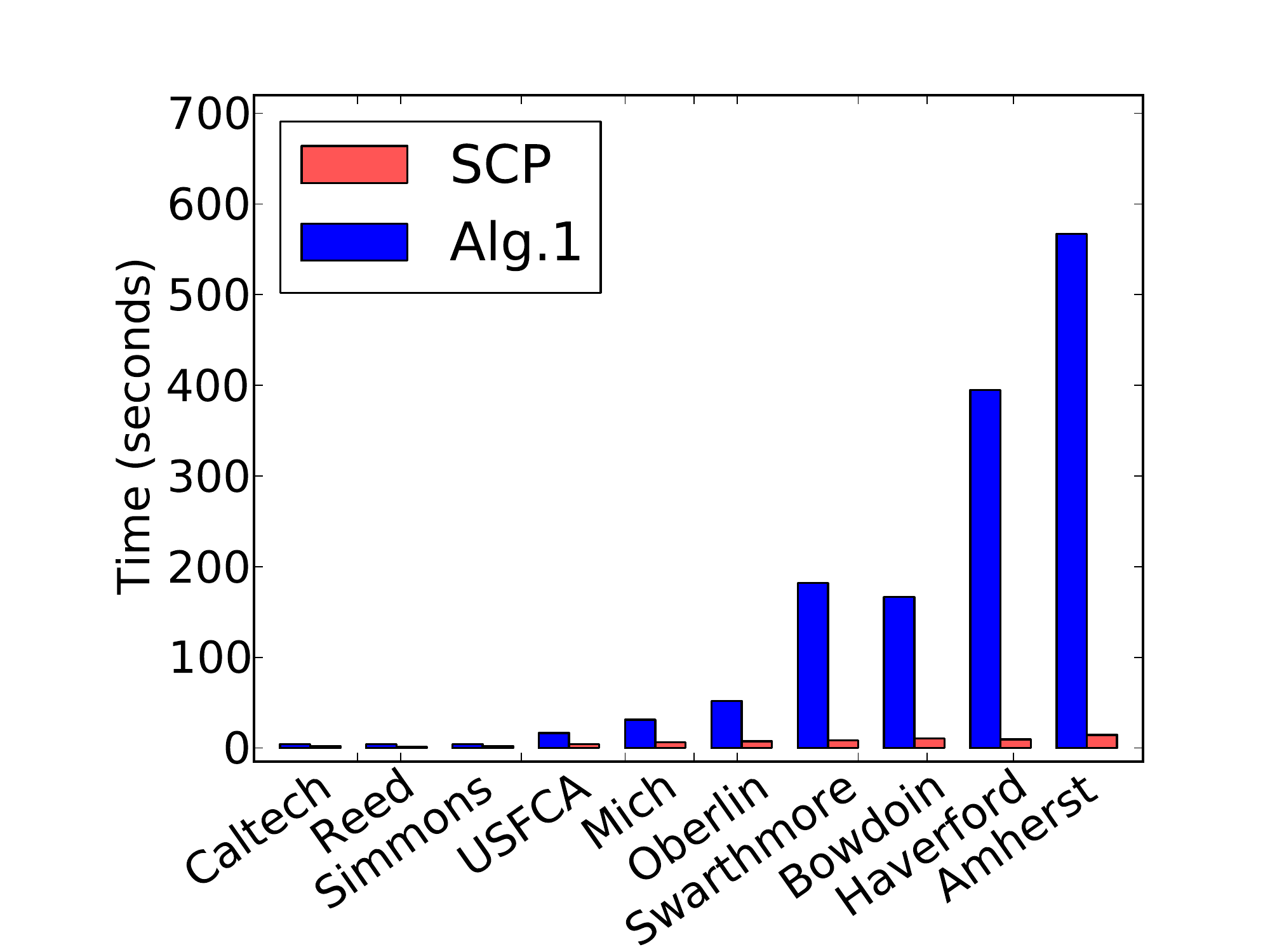}
        \label{K4}
    }
    \subfigure[K=5]{
        \includegraphics[width=41mm]{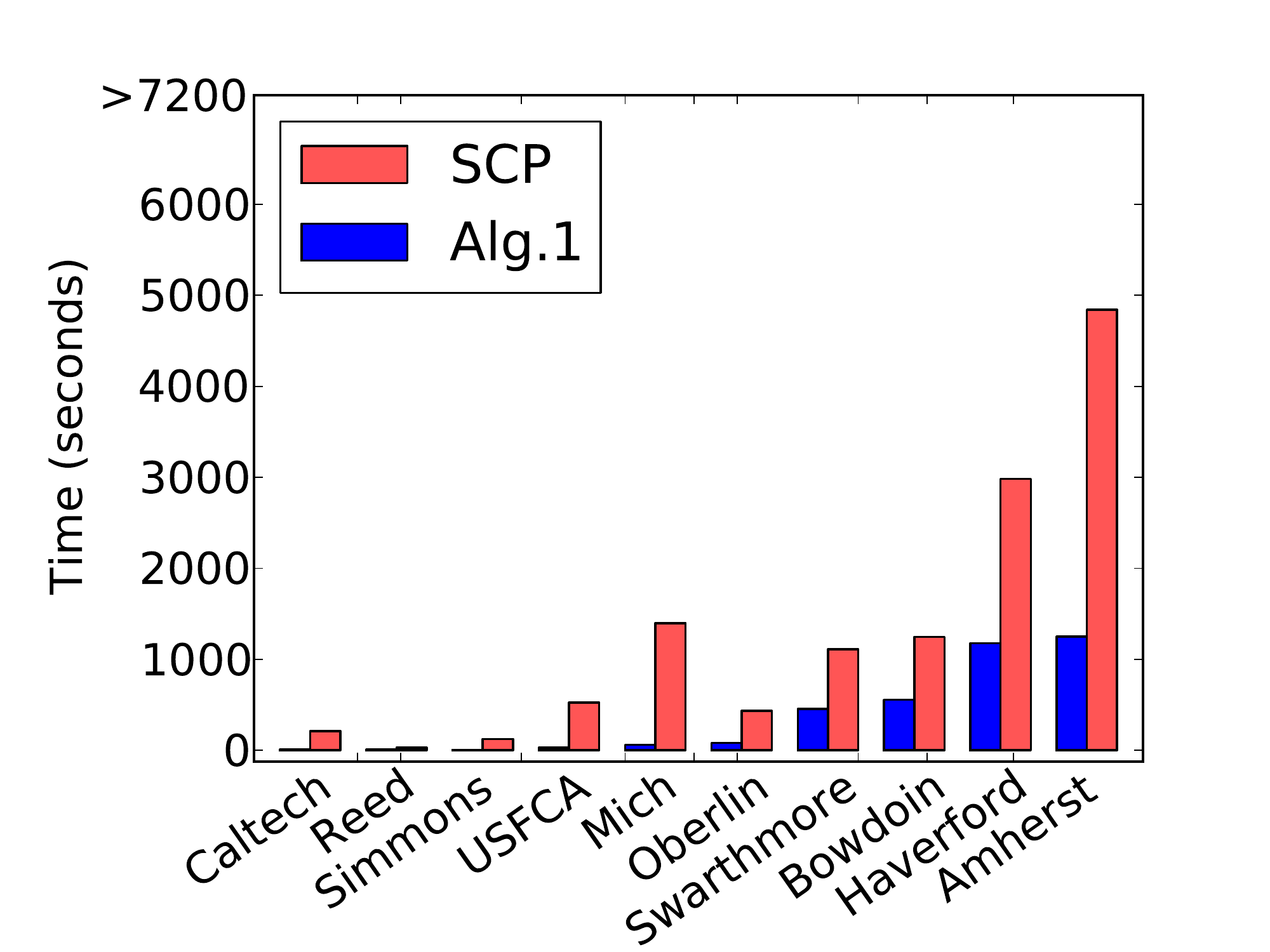}
        \label{K5}
    }
    \subfigure[K=6]{
        \includegraphics[width=41mm]{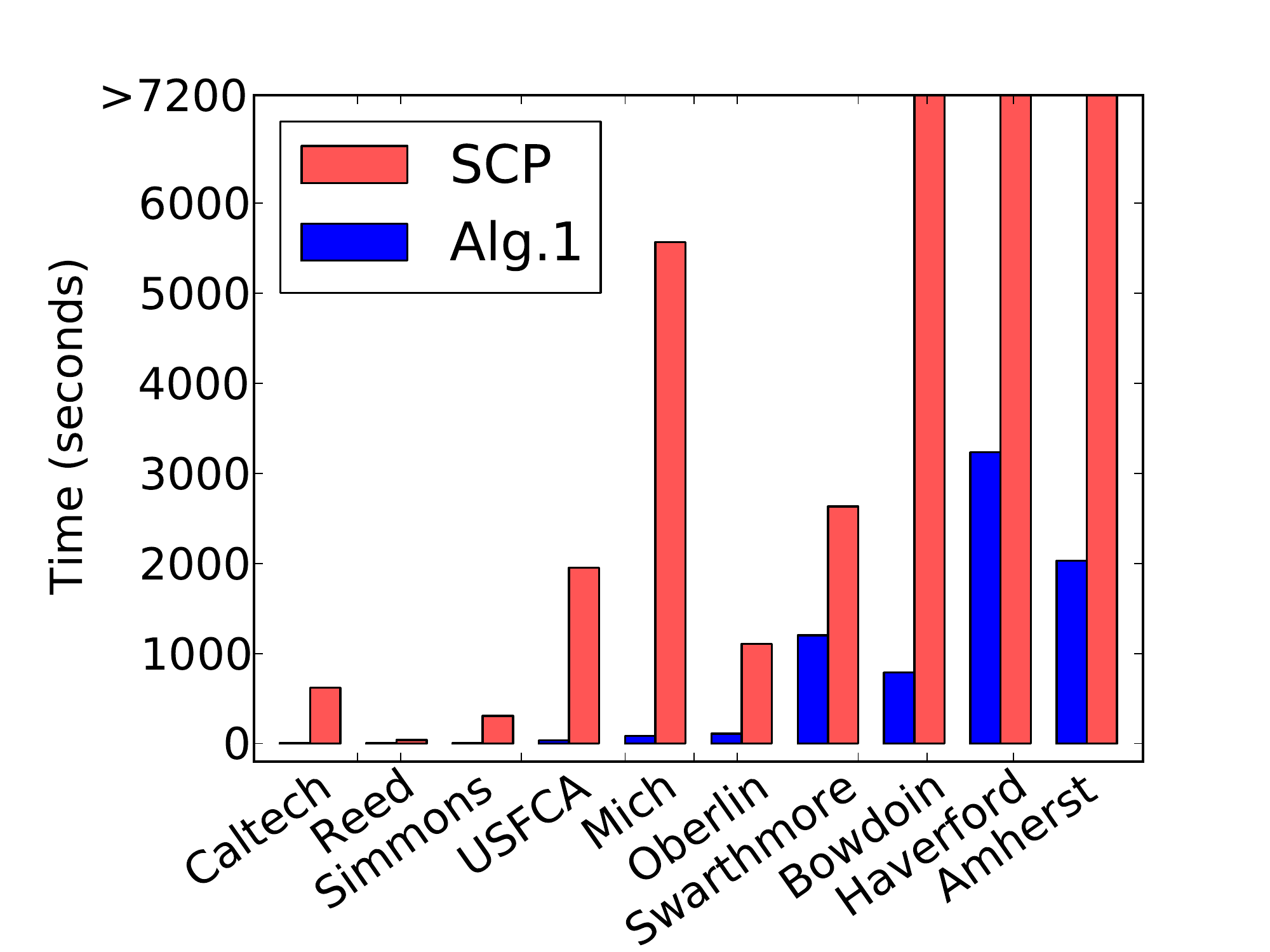}
        \label{K6}
    }
    \subfigure[K=7]{
        \includegraphics[width=41mm]{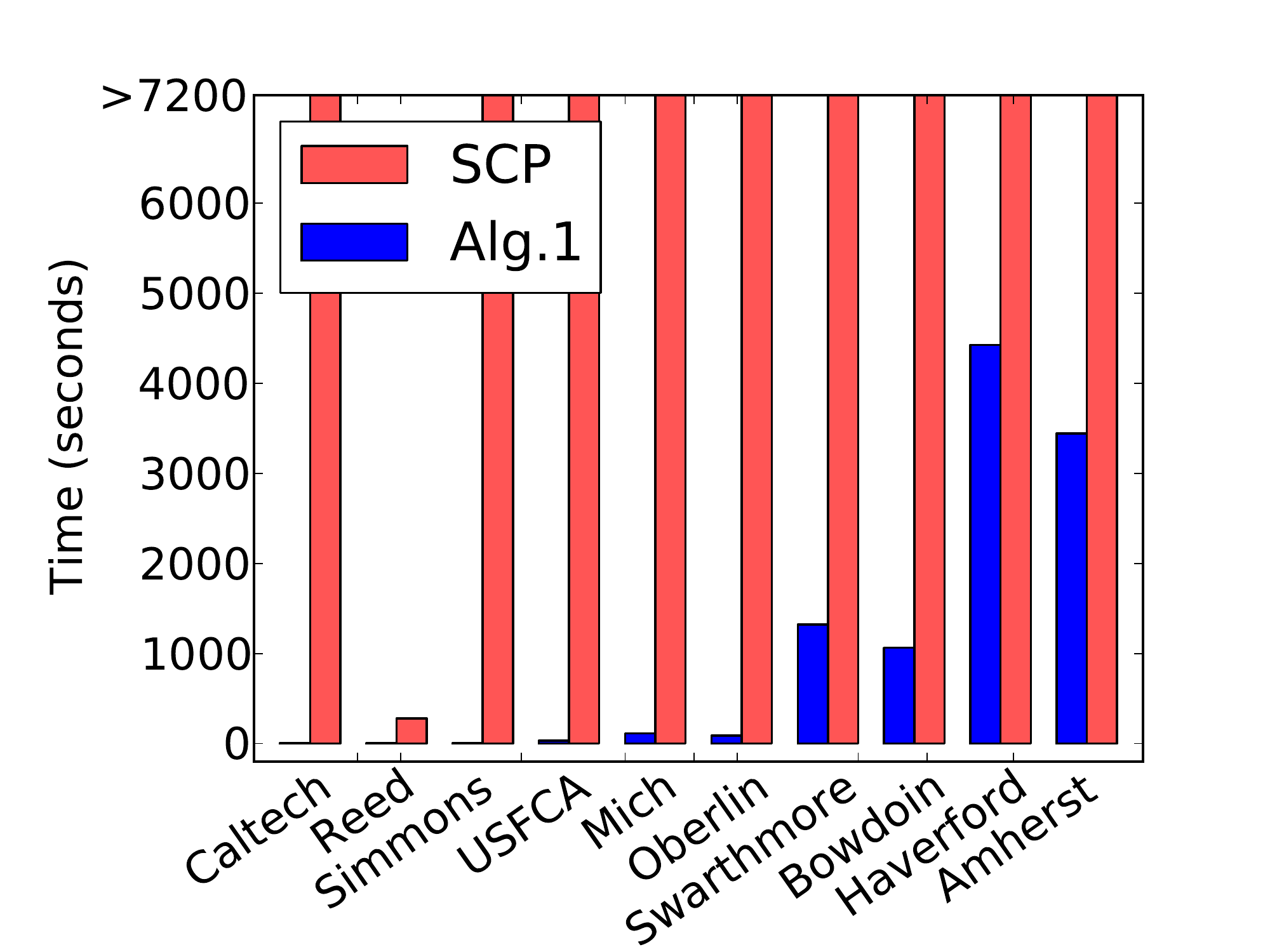}
        \label{K7}
    }
\caption{Benchmark results on the smallest 10 Facebook 100 graphs, for values of $k$ between $4$ and $7$. Red columns denote SCP runtimes, blue denote Algorithm 1. Processes were terminated if they failed to complete within 2 hours (7200 seconds). Running SCP for additional time sometimes resulted in it using more than 50GB of RAM; the SCP binary implementation consumes a vast amount of RAM for some networks and values of $k$, presumably for its disjoint-set forest fast intersection data structure. Our method never uses more than a small multiple of the amount used to store the maximal cliques; never more than ~512MB in these experiments. Experiments are run on a multiprocessor machine, in random sequence. SCP outperforms our algorithm for $k=4$, but performs substantially worse, in both time and space requirements, as values of $k$ get larger.}
\label{FacebookBenchmarkResults}
\end{center}
\vspace{-10px}
\end{figure}
}

\subsection{Counting the failed intersection tests}
This type of algorithm involves testing pairs of cliques that share a node, to see if they
intersect by at least $k$-1 nodes. For speed, our goal is to minimize the number of tests
required, as this typically takes up the major portion of the runtime
of such an algorithm.
Each test either succeeds, if there are $>=$ ($k$-1) nodes in the intersection,
or else fails.
We cannot change the minimum number of successful intersection tests --
there will be one such test for each edge in the minimal spanning tree,
i.e. $numCliques - numCommunities$.
But we can try to minimize the number of failed tests. In this subsection,
we will look at a lower bound on the number of failed tests that must
occur in Algorithm 1 described above.
In many networks, the number of these failed tests is very large
in comparison to the number of the successful tests, and dominates the computational cost.
This motivates our second algorithm, which we describe in the
next section.

Consider every pair of cliques that share at least one node.
We do not need to perform an intersection
test for every such pair -- we can potentially avoid testing most pairs of cliques which percolate into
the same community.
But we cannot avoid testing the pairs of cliques which share a node, but never percolate
into the same community.
For each network, and for each value of $k$, we count the 
pairs of cliques which share at least one node and which do not percolate, and divide this by the necessary number of successful tests.
Some of the highest ratios are shown in Table \ref{TBLratios}.
Even though Algorithm 1 is much faster
than SCP for higher $k$, this is a computational bottleneck. Each failed intersection test is
akin to a `false positive'; we tested the cliques that shared a node, in case they shared $k$-1, but this turned out not to be the case.  
\ifthenelse {\boolean{longVersion}} {
We aim to minimize the `false positive' rate -- the pairs of cliques that our heuristic `suspects' might percolate together, and hence need to be expensively tested for overlap against each other -- while of course ensuring
we still pick up all the true positives -- the cliques that actually do intersect by $k$-1 nodes.
}

\ifthenelse {\boolean{longVersion}} {
\begin{table}
	\centering
	\begin{tabular}{|l|r|r|r|r|}
		\hline
		University & k & Failed tests & Successful tests & Ratio \\
		\hline
		\hline
Berkeley13      & 4     &  \numprint{10466831}      & \numprint{1888501}       & 5.5424\\
UC33    & 5     & \numprint{36722163}      & \numprint{1396585}       & 26.294\\
JMU79   & 6     & \numprint{87790729}      & \numprint{1007912}       & 87.101\\
JMU79   & 7     & \numprint{202207552}     & \numprint{942903}        & 214.45\\
Baylor93        & 8     & \numprint{7979454075}    & \numprint{5667415}       & 1407.9\\
JMU79   & 9     & \numprint{2639954200}    & \numprint{852264}        & 3097.5\\
JMU79   & 10    & \numprint{4148901179}    & \numprint{817621}        & 5074.3\\
UC61    & 11    & \numprint{5257215707}    & \numprint{948764}        & 5541.1\\
Bingham82       & 12    & \numprint{3490950418}    & \numprint{511790}        & 6821.0\\
UC61    & 13    & \numprint{6271564288}    & \numprint{849242}        & 7384.8\\
Rutgers89       & 14    & \numprint{2988388319}    & \numprint{386558}        & 7730.7\\
Rutgers89       & 15    & \numprint{2388357546}    & \numprint{339698}        & 7030.8\\
MIT8    & 16    & \numprint{3941888908}    & \numprint{322212}        & 12233.\\
MIT8    & 17    & \numprint{3183156718}    & \numprint{285917}        & 11133.\\
Howard90        & 18    & \numprint{3675540181}    & \numprint{148661}        & 24724.\\
Howard90        & 19    & \numprint{3235463119}    & \numprint{138158}        & 23418.\\
UC61    & 20    & \numprint{15499684499}   & \numprint{582658}        & 26601.\\
Howard90        & 21    & \numprint{2770597042}    & \numprint{125715}        & 22038.\\
Howard90        & 22    & \numprint{2595044785}    & \numprint{121188}        & 21413.\\
Howard90        & 23    & \numprint{2443483517}    & \numprint{117057}        & 20874.\\
Howard90        & 24    & \numprint{2278591525}    & \numprint{112454}        & 20262.\\
Howard90        & 25    & \numprint{2090574938}    & \numprint{107072}        & 19524.\\
Howard90        & 26    & \numprint{1893432720}    & \numprint{101098}        & 18728.\\
Howard90        & 27    & \numprint{1644566572}    & \numprint{92900} & 17702.\\
Howard90        & 28    & \numprint{1356613382}    & \numprint{82661} & 16411.\\
Howard90        & 29    & \numprint{957002789}     & \numprint{67611} & 14154.\\
Howard90        & 30    & \numprint{588629576}     & \numprint{53215} & 11061.\\
Howard90        & 31    & \numprint{312582980}     & \numprint{42017} & 7439.4\\
Howard90        & 32    & \numprint{111014036}     & \numprint{33669} & 3297.2\\
Wake73  & 33    & \numprint{558217752}     & \numprint{136027}        & 4103.7\\
Wake73  & 34    & \numprint{243637026}     & \numprint{99631} & 2445.3\\
Tulane29        & 35    & \numprint{15603680}      & \numprint{8202}  & 1902.4\\
Wake73  & 36    & \numprint{33370824}      & \numprint{54029} & 617.64\\
Wake73  & 37    & \numprint{6927976}       & \numprint{38747} & 178.80\\
FSU53   & 38    & \numprint{1081780785}    & \numprint{4032339}       & 268.27\\
FSU53   & 39    & \numprint{231651220}     & \numprint{3728247}       & 62.134\\
FSU53   & 40    & \numprint{73496336}      & \numprint{3396986}       & 21.635\\
		\hline
\end{tabular}
\caption{Using the Facebook100 networks. For each $K$ between 4 and 20, the university with the highest ratio of failed intersection tests to the number of successful intersection tests is shown.
}
\label{TBLratios}
\vspace{-16px}
\end{table}
}
{
\begin{table}
	\centering
	\begin{tabular}{|l|r|r|r|r|}
		\hline
		University & k & Failed tests & Successful tests & Ratio \\
		\hline
		\hline
Berkeley13      & 4     &  \numprint{10466831}      & \numprint{1888501}       & 5.5424\\
UC33    & 5     & \numprint{36722163}      & \numprint{1396585}       & 26.294\\
JMU79   & 6     & \numprint{87790729}      & \numprint{1007912}       & 87.101\\
JMU79   & 7     & \numprint{202207552}     & \numprint{942903}        & 214.45\\
Baylor93        & 8     & \numprint{7979454075}    & \numprint{5667415}       & 1407.9\\
JMU79   & 9     & \numprint{2639954200}    & \numprint{852264}        & 3097.5\\
JMU79   & 10    & \numprint{4148901179}    & \numprint{817621}        & 5074.3\\
UC61    & 11    & \numprint{5257215707}    & \numprint{948764}        & 5541.1\\
Bingham82       & 12    & \numprint{3490950418}    & \numprint{511790}        & 6821.0\\
UC61    & 13    & \numprint{6271564288}    & \numprint{849242}        & 7384.8\\
Rutgers89       & 14    & \numprint{2988388319}    & \numprint{386558}        & 7730.7\\
Rutgers89       & 15    & \numprint{2388357546}    & \numprint{339698}        & 7030.8\\
MIT8    & 16    & \numprint{3941888908}    & \numprint{322212}        & 12233.\\
MIT8    & 17    & \numprint{3183156718}    & \numprint{285917}        & 11133.\\
Howard90        & 18    & \numprint{3675540181}    & \numprint{148661}        & 24724.\\
Howard90        & 19    & \numprint{3235463119}    & \numprint{138158}        & 23418.\\
UC61    & 20    & \numprint{15499684499}   & \numprint{582658}        & 26601.\\
		\hline
\end{tabular}
\caption{Using the Facebook100 networks. For each $K$ between 4 and 20, the university with the highest ratio of failed intersection tests to the number of successful intersection tests is shown.
}
\label{TBLratios}
\vspace{-16px}
\end{table}
}

\subsection{Algorithm 2}
Our second algorithm uses the same framework described in Section \ref{SECalgorithmTemplate}.
The only change is that we have a faster way, given the current frontier clique, of
finding its unvisited neighbouring cliques: a cheaper way of avoiding intersection tests, by detecting when two cliques will percolate in way that gives fewer `false positives' than checking only whether they share a single node.


We use a complete binary tree where there are as many leaf nodes
as there are maximal cliques.
We will refer to the nodes in
this binary tree as \emph{tree-nodes}, and the tree-nodes which are at
the leaves of our tree will be referred to as \emph{leaf-nodes}.
To avoid confusion, we use
\emph{graph-node} to refer to the nodes in our original network.
Each tree-node will have a set of graph-nodes associated with it.
For the leaf-nodes, this will be the set of graph-nodes that are
in the maximal clique which we associated with that leaf-node.
For the other tree-nodes in the binary tree, the set of graph-nodes
will be the union of the graph-nodes associated with the leaf-nodes descended from it.

This tree is built once at the start of the algorithm and 
does not change. But each leaf-node has a boolean field
associated with it which records whether the associated clique
has already been percolated into a community (i.e. it has been \emph{visited}).
These \emph{visited} fields will be initialized to False at the beginning, and 
will all gradually be set True as the algorithm proceeds and visits the cliques.
For the other tree-nodes, this boolean field will be set to True
when the fields of its children have been set to True.
This means that a boolean field will be True if and only if
all of its descendants have already been visited.


Given the current frontier clique, we could visit all the leaf nodes
and perform an intersection test against each.
This would be the naive algorithm again.
But we can ignore cliques that have been visited already.
In particular, if a tree-node has been marked as visited, then we can ignore all the cliques
that are descended from it.
Also, given a particular tree-node we can test the intersection between the current frontier
clique and the set of graph-nodes which are stored at this tree-node.
If this intersection is less than $k$-1, then we know that there cannot be a neighbouring
clique among the cliques descended from this tree-node.
These two observations allow us to skip most of the intersection tests via the
information present in the binary tree.

The search is a recursive search which starts at the root-node of the binary tree.
\ifthenelse {\boolean{longVersion}} {
Listing \ref{aaronsPseudocode} contains pseudo-code illustrating 
this algorithm.
}

At each tree-node, we check if this tree-node has been marked as visited,
and also check the size of the intersection. If the intersection is large enough
and if the tree-node is still unvisited, then the search will proceed
recursively into each of the two children of the tree-node.
Otherwise, the recursive function will return and that tree-node (and all its descendants)
will not be considered any further.

To save memory, we do not explicitly maintain a set of graph-nodes
at each non-leaf tree-node.
Instead we use a Bloom filter -- a fast and memory-efficient structure which probabilistically records
whether a given element is a member of a set.
It sometimes produces false positives, which
in our algorithm would sometimes lead to an overestimate
of the size of the intersection between two cliques; this does not cause a correctness problem, but does mean that sometimes the recursive search
proceeds slightly deeper than necessary.

\begin{lstlisting}[label={aaronsPseudocode},float=*,caption={Python-like pseudo-code description of our more complex clique percolation.}]
def get_unvisited_adjacent_cliques(current_clique, binary_tree, frontier):
	binary_search_in_tree(current_clique, binary_tree.root_node(), frontier)

def binary_search_in_tree(current_clique, tree_node, frontier):
	if not tree_node.visited_flag:
        if len(current_clique.intersection(tree.my_set_of_graph_nodes)) >= k-1:
            # tests pass, the child nodes are still candidates
            if tree_node.is_a_leaf_node():
                frontier.append(tree_node)
                tree_node.visited_flag = True
                tree_node.check_if_sibling_has_been_visited_and_propagate()
            else:
                binary_search_in_tree(current_clique, tree_node.left_child)
                binary_search_in_tree(current_clique, tree_node.right_child)
	else:
		# the children cannot be suitable, don't search the descendants
\end{lstlisting}

\ifthenelse {\boolean{longVersion}} {
\begin{table}
	\centering
	\begin{tabular}{|l|r|r|r|r|}
		\hline
		University & Nodes & Edges & Time(s) \\
		\hline
		\hline
Caltech36  &  \numprint{769} & \numprint{16656}  & 50.94  \\
Reed98  &  \numprint{962} & \numprint{18812}  & 51.84  \\
Simmons81  &  \numprint{1518} & \numprint{32988}  & 115.10  \\
Haverford76  &  \numprint{1446} & \numprint{59589}  & 372.39  \\
Swarthmore42  &  \numprint{1659} & \numprint{61050}  & 235.93  \\
USFCA72  &  \numprint{2682} & \numprint{65252}  & 215.22  \\
Mich67  &  \numprint{3748} & \numprint{81903}  & 421.19  \\
Bowdoin47  &  \numprint{2252} & \numprint{84387}  & 246.39  \\
Oberlin44  &  \numprint{2920} & \numprint{89912}  & 263.45  \\
Amherst41  &  \numprint{2235} & \numprint{90954}  & 424.96  \\
Wellesley22  &  \numprint{2970} & \numprint{94899}  & 338.81  \\
Hamilton46  &  \numprint{2314} & \numprint{96394}  & 389.64  \\
Smith60  &  \numprint{2970} & \numprint{97133}  & 421.62  \\
Trinity100  &  \numprint{2613} & \numprint{111996}  & 477.63  \\
Williams40  &  \numprint{2790} & \numprint{112986}  & 524.29  \\
Vassar85  &  \numprint{3068} & \numprint{119161}  & 376.70  \\
Middlebury45  &  \numprint{3075} & \numprint{124610}  & 500.21  \\
Brandeis99  &  \numprint{3898} & \numprint{137567}  & 411.83  \\
Wesleyan43  &  \numprint{3593} & \numprint{138035}  & 622.84  \\
Santa74  &  \numprint{3578} & \numprint{151747}  & 840.05  \\
Pepperdine86  &  \numprint{3445} & \numprint{152007}  & 1454.71  \\
Colgate88  &  \numprint{3482} & 1\numprint{55043}  & 844.91  \\
UC64  &  \numprint{6833} & \numprint{155332}  & 967.06  \\
Bucknell39  &  \numprint{3826} & \numprint{158864}  & 788.16  \\
Rochester38  &  \numprint{4563} & \numprint{161404}  & 1158.68  \\
Rice31  &  \numprint{4087} & \numprint{184828}  & 1289.66  \\
JohnsHopkins55  &  \numprint{5180} & \numprint{186586}  & 1047.59  \\
Vermont70  &  \numprint{7324} & \numprint{191221}  & 751.05  \\
Lehigh96  &  \numprint{5075} & \numprint{198347}  & 921.42  \\
Howard90  &  \numprint{4047} & \numprint{204850}  & 2395.45  \\
\hline
\end{tabular}
\caption{The runtimes of our more optimized algorithm, to compute all values of $k$, on the smallest 10 Facebook networks, and a selection of larger ones. 
This algorithm was often able to complete all values of $k$ in less time than Algorithm 1 took to complete $k=4$.
For every dataset tested, this algorithm calculates any value of $k$ faster than Algorithm 1.
}
\label{BloomRuntimes}
\end{table}
}
{
\begin{table}
	\centering
	\begin{tabular}{|l|r|r|r|r|}
		\hline
		University & Nodes & Edges & Time(s) \\
		\hline
		\hline
Caltech36  &  \numprint{769} & \numprint{16656}  & 50.94  \\
Reed98  &  \numprint{962} & \numprint{18812}  & 51.84  \\
Simmons81  &  \numprint{1518} & \numprint{32988}  & 115.10  \\
Haverford76  &  \numprint{1446} & \numprint{59589}  & 372.39  \\
Swarthmore42  &  \numprint{1659} & \numprint{61050}  & 235.93  \\
USFCA72  &  \numprint{2682} & \numprint{65252}  & 215.22  \\
Mich67  &  \numprint{3748} & \numprint{81903}  & 421.19  \\
Bowdoin47  &  \numprint{2252} & \numprint{84387}  & 246.39  \\
Oberlin44  &  \numprint{2920} & \numprint{89912}  & 263.45  \\
Amherst41  &  \numprint{2235} & \numprint{90954}  & 424.96  \\
UMass92  &  \numprint{16516} & \numprint{519385}  & 18,544.87  \\
UConn91  &  \numprint{17212} & \numprint{604870}  & 27,875.35  \\
UPenn7  &  \numprint{14916} & \numprint{686501}  & 26,413.13  \\
UNC28  &  \numprint{18163} & \numprint{766800}  & 87,315.59  \\
USC35  &  \numprint{17444} & \numprint{801853}  & 84,481.39  \\

\hline
\end{tabular}
\caption{The runtimes of our more optimized algorithm, to compute all values of $k$, on the smallest 10 Facebook networks, and 5 larger ones. 
This algorithm was often able to complete all values of $k$ in less time than Algorithm 1 took to complete $k=4$.
For every dataset tested, this algorithm calculates any value of $k$ faster than Algorithm 1.
}
\vspace{-18px}
\label{BloomRuntimes}
\end{table}
}

\ifthenelse {\boolean{longVersion}} {
\subsection{Evaluation of Algorithm 2}
}

When considering a single value of $k$, Algorithm 2 was always faster than Algorithm 1,
except sometimes when $k=3$.
In Table \ref{BloomRuntimes}, we see the runtimes of Algorithm 2 on a number of Facebook networks.
These runtimes are the time taken to compute the communities for \emph{all} values of $k$.
This algorithm frequently computes all values of $k$ quicker than Algorithm 1 
can compute the communities for $k=4$.

\ifthenelse {\boolean{longVersion}} {

\section{Other Approaches}
\label{SECotherApproaches}

We now briefly mention other approaches to this problem that we have considered,  as well as other overlapping community finding algorithms of interest.

\subsection{Hierarchical Partitioning}
The work of Narasimhamurthy et al. \cite{narasimhamurthy2010partitioning} investigates speeding up $k$-clique percolation by dividing the network into partitions using a graph partitioning algorithm.
CFinder is then run on each partition, generating a set of $k$-clique communities for each partition.
The union of these sets of percolated communities is reported as the set of $k$-clique communities.
To evaluate their approach, they compare the set of communities found on the partitioned network against the communities found by running CFinder on the whole unpartitioned network.

This method fails to account for the case where $k$-clique communities span the border of the graph partitions.
This may not be a large problem in networks where communities -- even $k$-clique communities -- do not overlap.
However, as shown in our previous work \cite{reid2011partitioning} and discussed in other work \cite{ahn2010link}, \cite{leskovec2008statistical}, it will not be possible, on many networks, to partition the network without splitting many communities; as such, this method is an approximation technique not suitable for networks with pervasive community overlap, such as modern on-line social networks.
We have tried adapt this approximate method into an exact method, inspired by kd-Trees \cite{bentley1975multidimensional}, where cliques that overlap partitions are explicitly handled, by pushing them up the partition hierarchy, and testing them for percolation against the cliques at the same level, or below them, in the partition tree; however, we find that this method does not work as well as the related methods described in this paper; it may be useful in other percolation contexts.

\subsection{Stochastic Approximation}
As there are many more cliques than nodes, we considered whether working with a random subset of the maximal cliques would yield similar community structure, at lower computational cost than working with all maximal cliques.
We investigated randomly sampling a subset of the maximal cliques, and performing percolation with this sampled subset.
However, results were poor; the overlapping NMI \cite{lancichinetti2009detecting} of the $k$-clique communities found by this approach, compared with the communities found using all cliques, decreased rapidly when even a small percentage of cliques ($<$ 10\%) were randomly omitted.
Perhaps a more sophisticated sampling approach will be found in future work; but simple approaches to stochastic approximation appear unsuitable for this problem.

\subsection{Other structure definitions}
Many other definitions of community structure exist.
While it was the goal of this work to focus on improving the performance of $k$-clique percolation, researchers have done work on other definitions of structure, such as $K$-plexes, $K$-cores, and $K$-trusses.  A highly scalable $K$-truss implementation in particular has been provided \cite{cohen2009graph}.
Types of percolation structure which lack the properties described in Figure \ref{cliqueProblem2}, so that sub-structures can easily be composed into higher level units discarding the sub-structure, avoid some of the computational problems of $k$-clique percolation.
However, these structures are fundamentally different than $k$-clique percolation, and a discussion of them should also consider the field of community finding as a whole, which is beyond the scope of this computational work.

\subsection{Other overlapping community finding methods}
We have focused on $k$-clique percolation in this work. However, many other modern overlapping community finding algorithms exist, many of them with good scaling properties, such as link partitioning methods \cite{evans2009line} \cite{ahn2010link}, relatively scalable methods which approximate statistical objectives \cite{mcdaid2010detecting} \cite{lancichinetti2011finding}, information theoretic approaches \cite{kim2011map}, methods explicitly designed to be scalable, such as label propagation \cite{gregory2010finding}, other clique-based methods more scalable than $k$-clique percolation \cite{shen2009detect} \cite{lee2010detecting} and many other methods; Xie et al. \cite{xie2011overlapping} provide a comparative analysis.

We have not investigated these algorithms in this computational work, focusing instead on the properties of $k$-clique percolation; but for many application domains the structures found by these often more scalable algorithms may be more suitable than percolated $k$-cliques.

\subsection{Better, as yet unknown, $k$-clique percolation methods}
In our work, we have focused on algorithms for $k$-clique percolation that either find all the cliques, or all the maximal cliques. 
We have not definitively ruled out the possibility of creating a better $k$-clique percolation algorithm which does not need to calculate this information, or which uses some other form of intermediate structure when percolating $k$-cliques, that somehow leads to more efficient clique intersection testing. 
This may be a potential area of future work, especially if $k$-clique percolation were to continue to be widely used, and not be superseded by more recent overlapping community finding algorithms.
} 

\section{Conclusion}
We have examined $k$-clique percolation as a specific example of a computationally challenging percolation problem.
We have shown that vast numbers of cliques exist in empirical social networks, and that $k$-clique percolation is a hard problem to implement well, due to the difficulty of producing intermediate representations of percolating structures.
From the large number of overlapping cliques and maximal cliques that we observe in empirical networks, and the consequent large number of edges in the clique graphs we have constructed, we conclude that while clique graphs are an attractive conceptual tool, their utility is limited in applications of social network analysis.
We have developed and thoroughly benchmarked a $k$-clique percolation algorithm that is conceptually simple, yet performs better than existing methods on many real world networks, especially when considering $k$-clique percolation with higher values of $k$.
This method is challenged by the large numbers of cliques which share at least one node, but do not percolate.
We introduce a second method which uses a more sophisticated data structure. 
With this method we can conduct clique percolation on larger pervasively overlapping networks than ever before.
However, these methods remain fundamentally limited by the necessity of testing cliques against other cliques with which they share \emph{some} nodes, but do not percolate.
Given the number of cliques, clique percolation remains computationally challenging; other overlapping community detection methods appear more promising, from a computational standpoint.
We provide software for researchers studying other percolation problems to leverage our results.

\ifthenelse {\boolean{longVersion}} {
\section{Acknowledgements}
This work is supported by Science Foundation Ireland under grant 08/SRC/I1407: Clique: Graph and Network Analysis Cluster.
}
{
}



\bibliographystyle{IEEEtran}
%

\bibliography{IEEEabrv,mybib}

\end{document}